\documentclass[12pt]{article}
\usepackage{a4wide}
\usepackage{axodraw}
\usepackage{latexsym}

\usepackage{pslatex}
\usepackage[latin1]{inputenc}
\usepackage[T1]{fontenc}

\def\bq{\begin{eqnarray}}
\def\eq{\end{eqnarray}}

\def\eps{\varepsilon}

\newlength{\dinwidth} \newlength{\dinmargin}
\begin{document}
\thispagestyle{empty}

\begin{flushright}
  TTP 01-25 \\
  UPRF-2001-21
\end{flushright}

\vspace{1.5cm}

\begin{center}
  {\Large\bf Nested Sums, Expansion of Transcendental Functions
             and Multi-Scale Multi-Loop Integrals\\}
  \vspace{1cm}
  {\large Sven Moch$^a$, Peter Uwer$^a$ and Stefan Weinzierl$^b$}\\
  \vspace{1cm}
  $^a${\small {\em Institut f{\"u}r Theoretische Teilchenphysik,
  Universit{\"a}t Karlsruhe}}\\
  {\small {\em 76128 Karlsruhe, Germany}}\\[2mm]
  $^b${\small {\em Dipartimento di Fisica, Universit\`a di Parma,\\
       INFN Gruppo Collegato di Parma, 43100 Parma, Italy}} \\
\end{center}

\vspace{2cm}

\begin{abstract}\noindent
  {%
    Expansion of higher transcendental functions in a small parameter  
    are needed in many areas of science. 
    For certain classes of functions 
    this can be achieved by algebraic means.
    These algebraic tools are based on nested sums
    and can be formulated as algorithms suitable 
    for an implementation on a computer.
    Examples, such as expansions of generalized hypergeometric 
    functions or Appell functions are discussed.
    As a further application, we give the general solution of a two-loop
    integral, the so-called C-topology, in terms of multiple nested sums. 
    In addition, we discuss some important properties of nested sums, 
    in particular we show that they satisfy a Hopf algebra.
   }
\end{abstract}

\vspace*{\fill}

\newpage

\reversemarginpar

\section{Introduction}
\label{sec:intro}

The expansion of higher transcendental functions \cite{Erdelyi,Slater}
is a common problem occuring in many areas of science.
It is of particular interest in particle physics in the calculation 
of higher order radiative corrections to scattering amplitudes.
There, higher transcendental functions occur frequently in formal solutions 
for specific loop integrals.
The necessary expansions of these functions are in general a 
highly non-trivial task. 
This is particularly true, if the expansions are required to a very high order.

In calculations of higher order radiative corrections  
classical polylogarithms \cite{lewin:book}, 
as well as Nielsen's generalized polylogarithms \cite{Nielsen} appear.
However, this set of functions will not suffice, if the number of 
loops grows, or if several different scales are involved in the 
problem. Several extensions of this class of functions to multiple polylogarithms 
have been discussed recently 
\cite{Goncharov} - \cite{Gehrmann:2000zt}
.

It is the aim of this paper to perform a systematic study of multiple 
nested sums appearing in the expansion of higher transcendental functions around 
integer values of their indices. 
To that end, we define so called $Z$-sums, study their algebraic properties  
and discuss their relation to the multiple polylogarithms introduced in the literature 
\cite{Goncharov} - \cite{Gehrmann:2000zt}
.
We give algorithms to solve these multiple nested sums to any order in the expansion 
parameter $\eps$ in terms of a given basis in $Z$-sums.
All algorithms can be readily implemented on a computer.
The $Z$-sums can be considered as certain generalizations of Euler-Zagier 
sums \cite{Euler,Zagier} or of harmonic 
sums 
\cite{Gonzalez-Arroyo:1979df} - \cite{Blumlein:1998if} 
involving multiple ratios of scales.
The latter are known in physics since the calculation of higher order 
Mellin moments of the deep-inelastic structure 
functions 
\cite{Gonzalez-Arroyo:1979df}, \cite{Kazakov:1988jk} - \cite{Vermaseren:2000we}
.

At the same time, our results allow us to investigate higher loop multi-scale 
integrals occuring for instance in pertubative corrections 
to four-particle scattering amplitudes.
These integrals have received great attention in recent years, 
mainly motivated by calculations of the next-to-next-to-leading order corrections 
to amplitudes for Bhabha scattering \cite{Bern:2000ie}, for $p p \rightarrow 2\;\mbox{jets}$ 
\cite{Bern:2000dn} - \cite{Glover:2001af}
, for $p p \rightarrow \gamma \gamma$ \cite{Bern:2001dg} 
and for light-by-light scattering~\cite{Bern:2001df}.

The relevant master integrals at two loops with four external legs have been 
calculated using a variety of techniques.
Analytic results were obtained either with the help of Mellin-Barnes 
representations 
\cite{Smirnov:1999gc,Tausk:1999vh} 
or with differential equations \cite{Gehrmann:2000zt,Gehrmann:1999as}.
Numerical results were obtained 
by a numerical evaluation of the coefficients
of the $\eps$-expansion
\cite{Binoth:2000ps,Laporta:2000dc}
.
Here, we want to advocate a different point of view based on multiple nested sums.
As a new result and to illustrate our approach, 
we discuss a specific two-loop integral, the so-called 
C-topology with one leg offshell, which can be reduced for arbitrary powers 
of the propagators and arbitrary dimensions to the aforementioned sums.
This is useful for the calculation of the two-loop amplitude for 
$e^+ e^- \rightarrow \;\mbox{3 jets}$.
Some of the techniques presented here have already been used in a recent calculation 
with massive fermions \cite{Phaf:2001gc}. 
In addition, there exists a wide variety of 
related literature on higher transcendental functions occuring in loop integrals 
and we can only mention a few of them here 
\cite{Davydychev:1993ut} - \cite{Tarasov:2000sf}
.

This papers is organized as follows. 
In the next section we introduce nested sums, show that they satisfy an algebra
and summarize some important special cases of our definitions.
Section~\ref{sec:algo} contains the main results of this paper, 
in particular the algorithms for solving certain classes of nested sums. 
In sec.~\ref{sec:applications} we give some examples for expansions of generalized 
hypergeometric functions, Appell functions and the Kamp\'e de F\'eriet function.
As an application to higher loop multi-scale integrals, we discuss the C-topology 
in sec.~\ref{subsec:Ctopo}.
In Appendix A we show that the algebraic structure of nested sums
forms a Hopf algebra~\cite{Hopf,Milnor}.
In Appendix B we briefly review the multiple polylogarithms of
Goncharov \cite{Goncharov}.

\section{Definition and properties of nested sums}
\label{sec:def}

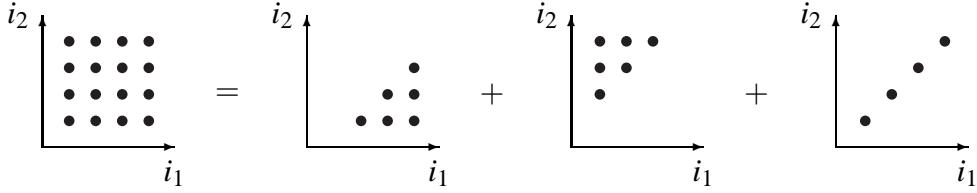
\begin{figure}
\begin{center}
\begin{picture}(400,100)(0,0)
\put(10,10){\vector(1,0){50}}
\put(10,10){\vector(0,1){50}}
\Text(60,5)[t]{$i_1$}
\Text(5,60)[r]{$i_2$}
\Vertex(20,20){2}
\Vertex(30,20){2}
\Vertex(40,20){2}
\Vertex(50,20){2}
\Vertex(20,30){2}
\Vertex(30,30){2}
\Vertex(40,30){2}
\Vertex(50,30){2}
\Vertex(20,40){2}
\Vertex(30,40){2}
\Vertex(40,40){2}
\Vertex(50,40){2}
\Vertex(20,50){2}
\Vertex(30,50){2}
\Vertex(40,50){2}
\Vertex(50,50){2}
\Text(80,30)[c]{$=$}
\put(110,10){\vector(1,0){50}}
\put(110,10){\vector(0,1){50}}
\Text(160,5)[t]{$i_1$}
\Text(105,60)[r]{$i_2$}
\Vertex(130,20){2}
\Vertex(140,20){2}
\Vertex(150,20){2}
\Vertex(140,30){2}
\Vertex(150,30){2}
\Vertex(150,40){2}
\Text(180,30)[c]{$+$}
\put(210,10){\vector(1,0){50}}
\put(210,10){\vector(0,1){50}}
\Text(260,5)[t]{$i_1$}
\Text(205,60)[r]{$i_2$}
\Vertex(220,30){2}
\Vertex(220,40){2}
\Vertex(230,40){2}
\Vertex(220,50){2}
\Vertex(230,50){2}
\Vertex(240,50){2}
\Text(280,30)[c]{$+$}
\put(310,10){\vector(1,0){50}}
\put(310,10){\vector(0,1){50}}
\Text(360,5)[t]{$i_1$}
\Text(305,60)[r]{$i_2$}
\Vertex(320,20){2}
\Vertex(330,30){2}
\Vertex(340,40){2}
\Vertex(350,50){2}
\end{picture}
\caption{\label{proof} Sketch of the proof for the multiplication of $Z$-sums. The sum over the square is replaced by
the sum over the three regions on the r.h.s.}
\end{center}
\end{figure}

We define the $Z$-sums by
\bq
\label{definition}
Z(n) & = & \left\{ \begin{array}{cc}
1, & n \ge 0, \\
0, & n < 0, \\
\end{array} \right.
\nonumber \\
Z(n;m_1,...,m_k;x_1,...,x_k) & = & \sum\limits_{i=1}^n \frac{x_1^i}{i^{m_1}} Z(i-1;m_2,...,m_k;x_2,...,x_k),
\eq
$k$ is called the depth, $w=m_1+...+m_k$ the weight.
An equivalent definition is given by
\bq 
  Z(n;m_1,...,m_k;x_1,...,x_k) & = & \sum\limits_{n\ge i_1>i_2>\ldots>i_k>0}
     \frac{x_1^{i_1}}{{i_1}^{m_1}}\ldots \frac{x_k^{i_k}}{{i_k}^{m_k}}.
\eq
In a similar way we define the $S$-sums by
\bq
\label{S-definition}
S(n) & = & \left\{ \begin{array}{cc}
1, & n > 0, \\
0, & n \le 0, \\
\end{array} \right.
\nonumber \\
S(n;m_1,...,m_k;x_1,...,x_k) & = & \sum\limits_{i=1}^n \frac{x_1^i}{i^{m_1}} S(i;m_2,...,m_k;x_2,...,x_k).
\eq
Once again an equivalent representation is given by:
\bq
S(n;m_1,...,m_k;x_1,...,x_k)  & = & 
\sum\limits_{n\ge i_1 \ge i_2\ge \ldots\ge i_k \ge 1}
\frac{x_1^{i_1}}{{i_1}^{m_1}}\ldots \frac{x_k^{i_k}}{{i_k}^{m_k}}.
\eq
The $S$-sums are closely related to the $Z$-sums, the difference being the upper summation boundary
for the nested sums: $(i-1)$ for $Z$-sums, $i$ for $S$-sums.
We introduce both $Z$-sums and $S$-sums, since some properties are more naturally expressed in terms of
$Z$-sums while others are more naturally expressed in terms of $S$-sums.
We can easily convert from the 
notation with $Z$-sums to the notation
with $S$-sums and vice versa:
\bq
\label{conversion}
S(n;m_1,...;x_1,...) & = & 
 \sum\limits_{i_1=1}^n \frac{x_1^{i_1}}{{i_1}^{m_1}} 
   \sum\limits_{i_2=1}^{i_1-1} \frac{x_2^{i_2}}{{i_2}^{m_2}}
   S(i_2;m_3,...;x_3,...) 
\nonumber \\ & &
 + S(n;m_1+m_2,m_3,...;x_1 x_2,x_3,...),
 \nonumber \\
Z(n;m_1,...;x_1,...) & = & 
 \sum\limits_{i_1=1}^n \frac{x_1^{i_1}}{{i_1}^{m_1}} 
   \sum\limits_{i_2=1}^{i_1} \frac{x_2^{i_2}}{{i_2}^{m_2}}
   Z(i_2-1;m_3,...;x_3,...) 
\nonumber \\ & &
 - Z(n;m_1+m_2,m_3,...;x_1 x_2,x_3,...).
\eq
The first formula allows to convert recursively a $S$-sum into a $Z$-sum.
The second
formula yields the conversion from a $Z$-sum to a $S$-sum.
For example in terms of $S$-sums the $Z$-sum
$Z(n;m_1,m_2,m_3,x_1,x_2,x_3)$ reads
\bq
Z(n;m_1,m_2,m_3,x_1,x_2,x_3) 
 & = &  
 S(n;m_1,m_2,m_3,x_1,x_2,x_3) 
 - S(n;m_1+m_2,m_3,x_1 x_2,x_3)
\nonumber \\
& &
 - S(n;m_1,m_2+m_3,x_1,x_2 x_3)
 + S(n;m_1+m_2+m_3,x_1 x_2 x_3).
\nonumber \\
\eq
Furthermore the $Z$-sums and the $S$-sums obey an algebra. 
A product of two $Z$-sums with the same upper summation limit can be written in terms
of single $Z$-sums. 
A straightforward generalization of the
results given by Vermaseren on the multiplication of harmonic sums
yields \cite{Vermaseren:1998uu}:
\bq
\label{Zmultiplication}
\lefteqn{
Z(n;m_1,...,m_k;x_1,...,x_k) \times Z(n;m_1',...,m_l';x_1',...,x_l') } & & \nonumber \\
& = & \sum\limits_{i_1=1}^n \frac{x_1^{i_1}}{i_1^{m_1}} Z(i_1-1;m_2,...,m_k;x_2,...,x_k) Z(i_1-1;m_1',...,m_l';x_1',...,x_l') \nonumber \\
&  & + \sum\limits_{i_2=1}^n \frac{x_1'^{i_2}}{i_2^{m_1'}} Z(i_2-1;m_1,...,m_k;x_1,...,x_k) Z(i_2-1;m_2',...,m_l';x_2',...,x_l') \nonumber \\
&  & + \sum\limits_{i=1}^n \frac{\left(x_1 x_1' \right)^{i}}{i^{m_1+m_1'}} Z(i-1;m_2,...,m_k;x_2,...,x_k) Z(i-1;m_2',...,m_l';x_2',...,x_l').
\eq
Recursive application of eq. (\ref{Zmultiplication}) leads to single $Z$-sums.
The proof of eq. (\ref{Zmultiplication}) 
follows immediately
from the relation 
\bq
  \sum_{i=1}^n\sum_{j=1}^n a_{ij} & = &
  \sum_{i=1}^n\sum_{j=1}^{i-1} a_{ij}
  +\sum_{j=1}^n\sum_{i=1}^{j-1} a_{ij}
  +\sum_{i=1}^n a_{ii},
\eq
which is sketched in fig. \ref{proof}.
Note that eq. (\ref{Zmultiplication})
directly translates into an algorithm for the multiplication of two $Z$-sums.
Details on the implementation of this algorithms on a computer 
can be found for example in \cite{Vermaseren:1998uu}.
We give an example for the product of two $Z$-sums:
\bq
Z(n;m_1,m_2;x_1,x_2) \times Z(n;m_3;x_3) 
 & = & 
 Z(n;m_1,m_2,m_3;x_1,x_2,x_3) 
+ Z(n;m_1,m_3,m_2;x_1,x_3,x_2) \nonumber \\
& & + Z(n;m_3,m_1,m_2;x_3,x_1,x_2) 
+ Z(n;m_1,m_2+m_3;x_1,x_2x_3) \nonumber \\
& & + Z(n;m_1+m_3,m_2;x_1 x_3,x_2).
\eq
Note that the product conserves the weight.
The $Z$-sums form actually a Hopf algebra. 
More details can be found in appendix \ref{sec:hopf}.
\\
\\
The $S$-sums also obey an algebra. The basic formula reads
\bq
\label{Smultiplication}
\lefteqn{
S(n;m_1,...,m_k;x_1,...,x_k) \times S(n;m_1',...,m_l';x_1',...,x_l') } & & \nonumber \\
& = & \sum\limits_{i_1=1}^n \frac{x_1^{i_1}}{i_1^{m_1}} S(i_1;m_2,...,m_k;x_2,...,x_k) S(i_1;m_1',...,m_l';x_1',...,x_l') \nonumber \\
&  & + \sum\limits_{i_2=1}^n \frac{x_1'^{i_2}}{i_2^{m_1'}} S(i_2;m_1,...,m_k;x_1,...,x_k) S(i_2;m_2',...,m_l';x_2',...,x_l') \nonumber \\
&  & - \sum\limits_{i=1}^n \frac{\left(x_1 x_1' \right)^{i}}{i^{m_1+m_1'}} S(i;m_2,...,m_k;x_2,...,x_k) S(i;m_2',...,m_l';x_2',...,x_l').
\eq
Note the minus sign in front of the last term compared to
the corresponding formula for $Z$-sums.

\subsection{Special cases}
\label{subsec:special}

\begin{figure}
\begin{center}
\begin{picture}(300,240)(0,0)
\Text(150,230)[c]{Z-sums}
\Text(50,180)[c]{multiple polylogs}
\Text(250,180)[c]{Euler-Zagier sums}
\Text(50,130)[c]{harmonic polylogs}
\Text(50,80)[c]{Nielsen polylogs}
\Text(50,30)[c]{classical polylogs}
\Text(250,130)[c]{multiple zeta values}
\ArrowLine(50,190)(130,220)
\ArrowLine(250,190)(170,220)
\ArrowLine(50,140)(50,170)
\ArrowLine(250,140)(250,170)
\ArrowLine(50,90)(50,120)
\ArrowLine(50,40)(50,70)
\ArrowLine(185,140)(100,170)
\end{picture}
\caption{\label{inheritance} The inheritance diagram for $Z$-sums shows the relations
between the various special cases.}
\end{center}
\end{figure}
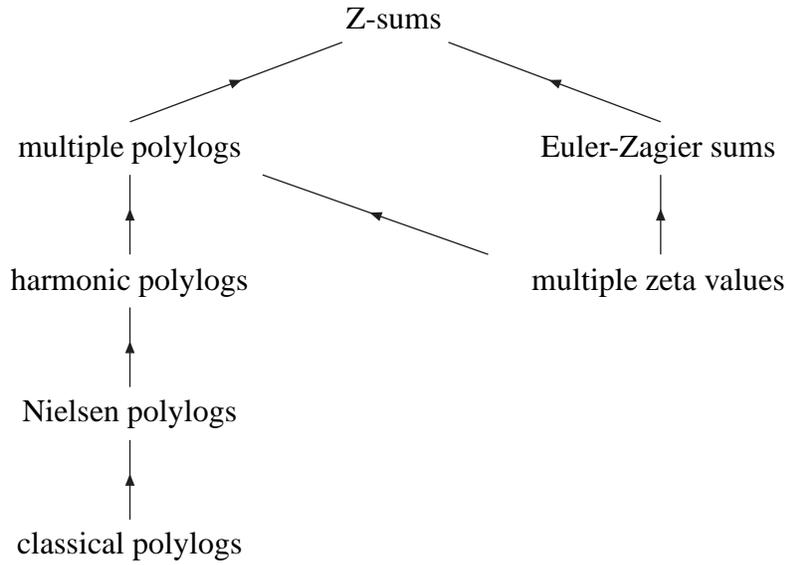

$Z$-sums and $S$-sums are generalizations of more known objects. 
We give here an overview of the most important special cases.\\
\\
For $n=\infty$ the $Z$-sums are the multiple polylogarithms of Goncharov \cite{Goncharov}:
\bq
\label{multipolylog}
Z(\infty;m_1,...,m_k;x_1,...,x_k) & = & \mbox{Li}_{m_k,...,m_1}(x_k,...,x_1).
\eq
For $x_1=...=x_k=1$ the definition reduces to the Euler-Zagier sums \cite{Euler,Zagier}:
\bq
Z(n;m_1,...,m_k;1,...,1) & = & Z_{m_1,...,m_k}(n).
\eq
For $n=\infty$ and $x_1=...=x_k=1$ the sum is a multiple $\zeta$-value \cite{Borwein}:
\bq
Z(\infty;m_1,...,m_k;1,...,1) & = & \zeta(m_k,...,m_1).
\eq
The $S$-sums reduce for $x_1=...=x_k=1$ (and positive $m_i$) to harmonic sums \cite{Vermaseren:1998uu}:
\bq
S(n;m_1,...,m_k;1,...,1) & = & S_{m_1,...,m_k}(n).
\eq
The multiple polylogarithms of Goncharov contain as the notation already suggests as subsets 
the classical polylogarithms 
$
\mbox{Li}_n(x)
$ 
\cite{lewin:book},
as well as
Nielsen's generalized polylogarithms \cite{Nielsen}
\bq
S_{n,p}(x) & = & \mbox{Li}_{1,...,1,n+1}(\underbrace{1,...,1}_{p-1},x),
\eq
the harmonic polylogarithms of Remiddi and Vermaseren \cite{Remiddi:1999ew}
\bq
\label{harmpolylog}
H_{m_1,...,m_k}(x) & = & \mbox{Li}_{m_k,...,m_1}(\underbrace{1,...,1}_{k-1},x)
\eq
and the two-dimensional harmonic polylogarithms introduced recently by
Gehrmann and Remiddi \cite{Gehrmann:2000zt}.
The exact connection to the two-dimensional harmonic polylogarithms 
is shown in appendix \ref{sec:polylog} together with a brief review
of the multiple polylogarithms of Goncharov.
Euler-Zagier sums and harmonic sums occur in the expansion of Gamma functions:
For positive integers $n$ we have 
on the positive side 
\bq
\label{expgamma}
\lefteqn{
\Gamma(n+\eps) = \Gamma(1+\eps) \Gamma(n) } & & \nonumber \\
& & \times \left( 1 + \eps Z_1(n-1) + \eps^2 Z_{11}(n-1) + \eps^3 Z_{111}(n-1) + ... 
+ \eps^{n-1} Z_{11...1}(n-1) \right). \nonumber \\
\eq
On the negative side (again $n>0$) we have
\bq
\lefteqn{
\Gamma(-n+1+\eps) } & & \nonumber \\
& = & \frac{\Gamma(1+\eps)}{\eps}
\frac{(-1)^{n-1}}{\Gamma(n)}
\left( 1 + \eps S_1(n-1) + \eps^2 S_{11}(n-1) + \eps^3 S_{111}(n-1) + ...
\right). \nonumber \\
\eq
The usefulness of the $Z$-sums lies in the fact, that they interpolate between
Goncharov's multiple polylogarithms and Euler-Zagier sums.
In addition, the interpolation is compatible with the algebra structure.
Fig. \ref{inheritance} summarizes the relations between the various special cases.

\section{Algorithms}
\label{sec:algo}

In this section we give the detailed algorithms which allow to solve 
the $\eps$-expansion of 
nested transcendental sums in terms of $Z$-sums or $S$-sums defined 
in eq.(\ref{definition}) and eq.(\ref{S-definition}), respectively.
By a transcendental sum we mean a sum over $i$ of finite or infinite summation range 
involving the following objects: 
\begin{enumerate}
\item Fractions of the form 
\bq
\frac{x^i}{(i+c)^m},
\eq
where $m$ is an integer, $c$ a non-negative integer and $x$ a real number.
\item Ratios of two Gamma functions
\bq
\label{ratio_tgamma}
\frac{\Gamma(i+a_1+b_1\eps)}{\Gamma(i+a_2+b_2\eps)},
\eq
where $a_1$ and $a_2$ are integers.
\item $Z$- and $S$-sums are also allowed to appear as subsums: 
\bq
     Z(i+c-1;m_1,...;x_1,...) & \mbox{or} & S(i+c';m_1,...;x_1,...).
\eq
The offsets $c$ and $c'$ are integers.
\item We further allow these building blocks to occur also with index $(n-i)$, for example
as in 
\bq
S(n-i;m_1',...;x_1',...).
\eq
Here $n$ denotes the upper summation limit.
\item In addition binomials 
\bq
       \left( \begin{array}{c} n \\ i \\ \end{array} \right)
 & = & \frac{n!}{i! (n-i)!}
\eq
may occur.
\end{enumerate}
Some examples of sums which can be constructed from these building blocks are
\bq
\sum\limits_{i=1}^n 
    \frac{x^i}{(i+4)^3} \frac{y^i}{(i+2)^8} 
    \frac{\Gamma(i+1+\eps)}{\Gamma(i+2+3\eps)}
& \mbox{and} & 
\sum\limits_{i=1}^{\infty} x^i \frac{\Gamma(i+a\eps)}{\Gamma(i+1-c\eps)}
      \frac{\Gamma(i+b\eps)}{\Gamma(i+1)}.
\eq
There are some simplifications, which can be done immediately:
Partial fractioning is used to reduce a product for $c_1 \neq c_2$
\bq
\frac{x_1^i}{(i+c_1)^{m_1}} \frac{x_2^i}{(i+c_2)^{m_2}} 
 & = &
        \frac{1}{c_2-c_1} \left[ \frac{x_1^i}{(i+c_1)^{m_1}} \frac{x_2^i}{(i+c_2)^{m_2-1}}
                                  - \frac{x_1^i}{(i+c_1)^{m_1-1}} \frac{x_2^i}{(i+c_2)^{m_2}} \right]\, ,
\nonumber \\
\eq
to terms which involve only the first factor or only the second one, but not both.\\
\\
Ratios of two Gamma functions as in eq. (\ref{ratio_tgamma}) are first reduced to the form
$\Gamma(i+b_1\eps)/\Gamma(i+b_2\eps)$ with the help of the identity
$\Gamma(x+1) = x \; \Gamma(x)$.
They are then expanded in $\eps$ using eq. (\ref{expgamma}).
To invert the power series which is obtained in the denominator the formula
\bq
\label{invertgammaseries}
\lefteqn{
\left( 1 + \eps Z_1(n-1) + \eps^2 Z_{11}(n-1) + \eps^3 Z_{111}(n-1) 
+ ... 
+ \eps^{n-1} Z_{11...1}(n-1) \right)^{-1} 
  } & & \nonumber \\ 
& = & 
1 - \eps S_1(n-1) + \eps^2 S_{11}(n-1) - \eps^3 S_{111}(n-1) + ...
\hspace*{5cm}
\eq
is useful to speed up the computation on a computer.\\
\\
There are also some basic operations involving $Z$- or $S$-sums.
First of all we can easily convert between the two notations, using eq. (\ref{conversion}).
Furthermore we would like to be able to relate the $Z$-sum $Z(n+c-1,...)$ to $Z(n-1,...)$ or
the $S$-sum $S(n+c,...)$ to $S(n,...)$, where $c>0$ is a fixed number.
This can easily be done with the help of the following formulae:
\bq
\label{adjust_downwards_Zsum}
\lefteqn{
     Z(n+c-1;m_1,...;x_1,...) 
} & & \nonumber \\
& = & Z(n-1;m_1,...;x_1,...) 
      + \sum\limits_{j=0}^{c-1} x_1^j \frac{x_1^n}{(n+j)^{m_1}} Z(n-1+j;m_2,...;x_2,...), \nonumber \\
\lefteqn{
     S(n+c;m_1,...;x_1,...) 
} & & \nonumber \\
& = & S(n;m_1,...;x_1,...) 
      + \sum\limits_{j=1}^{c} x_1^{j} \frac{x_1^n}{(n+j)^{m_1}} S(n+j;m_2,...;x_2,...).
\eq
The $Z$- or $S$-sums appearing in the last term have a reduced depth 
and the problem can be solved recursively.\\
Another situation which appears quite often is the product of two sums. 
If the upper summation limits of the two sums differ by some integer $c$ we first synchronize them with
the help of eq. (\ref{adjust_downwards_Zsum}).
For sums with equal upper summation limit 
one may use the 
algebra eq. (\ref{Zmultiplication}) or eq. (\ref{Smultiplication}) to convert 
the product into single sums of higher weight.\\
Furthermore we can bring $Z$-sums and $S$-sums to a standard form by eliminating letters
with negative degrees, that is positive powers of $i$.
In general, these cases are easy to handle. We illustrate this for $S$-sums.
We consider $S(n;-m_1,m_2,...;x_1,x_2,...)$, 
write out the outermost sum of the $S$-function 
and then interchange the order of summation:
\bq
\label{standard1}
S(n;-m_1,m_2,...;x_1,x_2,...)
& = & \sum\limits_{i_2=1}^n \frac{x_2^{i_2}}{i_2^{m_2}} S(i_2;m_3,...) 
 \sum\limits_{i_1=i_2}^n i_1^{m_1} x_1^{i_1}.
\eq
The inner sum can be evaluated for any given weight analytically. 
Subsequently the outer sum can be done with eq. (\ref{S-definition}).
If a negative weight occurs inside a sum, eq. (\ref{standard1}) is applied
to the subsum starting from the negative weight.
\\
\\
If a binomial appears in the sum, this sum may be written as a conjugation.
To any function $f(n)$ of an integer variable $n$ one defines the conjugated function $ C \circ f(n)$
as the following sum \cite{Vermaseren:1998uu}
\bq
C \circ f(n) & = & - \sum\limits_{i=1}^n \left( \begin{array}{c} n \\ i \\ \end{array} \right) (-1)^i f(i).
\eq 
Conjugation satisfies the following two properties
\bq
C \circ 1 & = & 1, \\
C \circ C \circ f(n) & = & f(n),
\eq
which can be easily verified.\\
\\
We classify four types of transcendental sums, 
which are dealt with in the algorithms A to D.
\begin{enumerate}
\item Sum over $i$ involving only $Z(i-1;...)$ (type A).
\item Sum over $i$ involving both $Z(i-1;...)$ and $Z(n-i-1;...)$ (type B).
\item Sum over $i$ involving $S(i;...)$ and a binomial (type C).
\item Sum over $i$ involving both $S(i;...)$ , $S(n-i;...)$ and a binomial (type D). 
\end{enumerate}
Many of the algorithms use a recursion. They relate a given problem to a simpler one,
either with a reduced depth or weight of the $Z$-sums or $S$-sums involved. 
In these cases we only give one step in the recursion. \\
The algorithms presented in this paper are 
all suited for programming in a computer algebra 
system like GiNaC~\cite{Bauer:2000cp}, FORM~\cite{Vermaseren:2000nd} or the commercial 
ones like Mathematica or Maple. 
Implementations within the GiNaC framework 
and in FORM along the lines of ref.~\cite{Vermaseren:1998uu}
are in preparation or  
have been published elsewhere \cite{gtybalt}.

\subsection{Algorithm A}
\label{subsubsec:type_A}

Here we consider sums of the form
\bq
     \sum\limits_{i=1}^n \frac{x^i}{(i+c)^m} 
       \frac{\Gamma(i+a_1+b_1\varepsilon)}{\Gamma(i+c_1+d_1\varepsilon)} ...
       \frac{\Gamma(i+a_k+b_k\varepsilon)}{\Gamma(i+c_k+d_k\varepsilon)}
       Z(i+o-1,m_1,...,m_l,x_1,...,x_l)
\eq
and show how to reduce them to $Z$-sums.
We assume that all $a_j$ and $c_j$ are integers, 
$c$ is assumed to be a nonnegative integer and $o$ should be an integer.
The upper summation limit is allowed to be infinity.
\\
\\
After expanding the Gamma functions and synchronizing the subsum
$Z(i+o-1,m_1,...)$
the problem is reduced to
sums of the form
\bq
     \sum\limits_{i=1}^n \frac{x^i}{(i+c)^m} Z(i-1,...)
\eq
with $c \ge 0$. 
It remains to reduce the offset $c$ to zero.
If the depth of the subsum is zero, we have
\bq
    \sum\limits_{i=1}^n \frac{x^i}{(i+c)^m}
     =
       \frac{1}{x} \sum\limits_{i=1}^n \frac{x^i}{(i+c-1)^m}
       \mbox{} - \frac{1}{c^m} + \frac{x^n}{(n+c)^m}.
\eq
The last term contributes only if $n$ is not equal to infinity.
If the depth of the subsum is not equal to zero, we have
\bq
\lefteqn{
     \sum\limits_{i=1}^n \frac{x^i}{(i+c)^m} Z(i-1,...)
= 
       \frac{1}{x} \sum\limits_{i=1}^n \frac{x^i}{(i+c-1)^m} Z(i-1,...)
} & & \nonumber \\
& &
       \mbox{} - \sum\limits_{i=1}^{n-1} \frac{x^i}{(i+c)^m} \frac{x_1^i}{i^{m_1}} Z(i-1,m_2,...)
       \mbox{} + \frac{x^n}{(n+c)^m} Z(n-1,...).
\eq
Note that the third term 
only contributes if $n$ is not equal to infinity.
Finally we arrive at
\bq
\label{result_type_A}
     \sum\limits_{i=1}^n \frac{x^i}{i^m} Z(i-1,...),
\eq
which is again a $Z$-sum. If the upper summation limit $n$ equals infinity this sum
yields immediately a multiple polylogarithm according to eq. (\ref{multipolylog}).
In the special case where $n$ equals infinity and the subsum is an Euler-Zagier sum
we obtain a harmonic polylogarithm according to eq. (\ref{harmpolylog}).

\subsection{Algorithm B}
\label{subsubsec:type_B}

Here we consider sums of the form
\bq
\lefteqn{
     \sum\limits_{i=1}^{n-1} 
       \frac{x^i}{(i+c)^m} 
       \frac{\Gamma(i+a_1+b_1\varepsilon)}{\Gamma(i+c_1+d_1\varepsilon)}
...
       \frac{\Gamma(i+a_k+b_k\varepsilon)}{\Gamma(i+c_k+d_k\varepsilon)} \, 
       Z(i+o-1,m_1,...,m_l,x_1,...,x_l) }
\nonumber \\
& &
       \times
       \frac{y^{n-i}}{(n-i+c')^{m'}} 
       \frac{\Gamma(n-i+a_1'+b_1'\varepsilon)}{\Gamma(n-i+c_1'+d_1'\varepsilon)}
...
       \frac{\Gamma(n-i+a_{k'}'+b_{k'}'\varepsilon)}{\Gamma(n-i+c_{k'}'+d_{k'}'\varepsilon)}
\hspace*{4.0cm}
\nonumber \\[1ex]
& & \qquad\qquad \times
       Z(n-i+o'-1,m_1',...,m_{l'}',x_1',...,x_{l'}')
\eq
and show how to reduce them to $Z$-sums.
Here, all $a_j$, $a_j'$, $c_j$  and $c_j'$ should be integers, $c$, $c'$
should be nonnegative integers and $o$,
$o'$ should be integers.
Note that the upper summation limit is $(n-1)$.
The upper summation limit should not be infinity.\\
\\
Using the expansion of the Gamma functions and the synchronization of the subsums,
we immediately obtain sums of the form
\bq
    \sum\limits_{i=1}^{n-1} \frac{x^i}{(i+c)^m} Z(i-1,m_1,...) \frac{y^{n-i}}{(n-i+c')^{m'}} Z(n-i-1,m_1',....).
\eq
Partial fractioning 
(and a change of the summation index $i \rightarrow n-i$ in sums involving
the fraction with $(n-i+c')$)
reduces these sums further to
sums of the type
\bq
\label{type_B_one_letter_only}
    \sum\limits_{i=1}^{n-1} \frac{x^i}{(i+c)^m} Z(i-1,m_1,...) Z(n-i-1,m_1',....).
\eq
If the depth of $Z(n-i-1,m_1',....)$ is zero, we have a sum of type A with upper summation index $(n-1)$:
\bq
    \sum\limits_{i=1}^{n-1} \frac{x^i}{(i+c)^m} Z(i-1,m_1,...).
\eq
Otherwise we can rewrite eq. (\ref{type_B_one_letter_only}) as
\bq
    \sum\limits_{j=1}^{n-1} 
    \left[ \sum\limits_{i=1}^{j-1} \frac{x^i}{(i+c)^m} Z(i-1,m_1,...) 
           \frac{{x_1'}^{j-i}}{(j-i)^{m_1'}} Z(j-i-1,m_2',....)
    \right]
\eq
and use recursion. The inner sum is again of type B, but with a reduced depth, 
such that the recursion will finally terminate.

\subsection{Algorithm C}
\label{subsubsec:type_C}

Here we consider sums of the form
\bq
\lefteqn{
       \mbox{} - \sum\limits_{i=1}^n 
       \left( \begin{array}{c} n \\ i \\ \end{array} \right)
       \left( -1 \right)^i
       \frac{x^i}{(i+c)^m} 
       \frac{\Gamma(i+a_1+b_1\varepsilon)}{\Gamma(i+c_1+d_1\varepsilon)}
       ...
       \frac{\Gamma(i+a_k+b_k\varepsilon)}{\Gamma(i+c_k+d_k\varepsilon)}
} & & \nonumber \\
& & \times
       S(i+o,m_1,...,m_l,x_1,...,x_l),
\hspace*{6cm}
\eq
where $a_j$ and $c_j$ are integers, $c$ 
is a nonnegative integer and $o $ is an integer.
The upper summation limit should not be infinity.
These sums cannot be reduced to $Z$-sums with upper summation limit $n$ alone.
However, they can be reduced to $Z$-sums with upper summation limit $n$ and
multiple polylogarithms (which are $Z$-sums to infinity).
\\
\\
Again, we expand the Gamma functions and synchronize the subsum.
It is therefore sufficient to consider sums of the form
\bq
     \mbox{} - \sum\limits_{i=1}^n 
       \left( \begin{array}{c} n \\ i \\ \end{array} \right)
       \left( -1 \right)^i
      \frac{x^i}{(i+c)^m} S(i,...)
\eq
with $c \ge 0$. 
To reduce the offset $c$ to zero we rewrite the sum as
\bq
    \left(-\frac{1}{x}\right) \frac{1}{n+1}
     (-1) \sum\limits_{i=1}^{n+1} 
       \left( \begin{array}{c} n+1 \\ i \\ \end{array} \right)
       \left( -1 \right)^i
      \frac{x^i}{(i+c-1)^m} i S(i-1,...).
\eq
Repeated application of the above relation yields 
sums of the form
\bq
\label{type_C_expanded}
     \mbox{} - \sum\limits_{i=1}^n 
       \left( \begin{array}{c} n \\ i \\ \end{array} \right)
       \left( -1 \right)^i
      \frac{x^i}{i^m} S(i,...).
\eq
If $m$ is negative we rewrite eq. (\ref{type_C_expanded})
as
\bq
     \mbox{} - n x (-1) \sum\limits_{i=1}^{n-1} 
       \left( \begin{array}{c} n-1 \\ i \\ \end{array} \right)
       \left( -1 \right)^i
      \frac{x^i}{(i+1)^{m+1}} S(i+1,...) + n x S(1,...).
\eq
We can therefore assume that $m$ is a non-negative number.
Furthermore, due to eq. (\ref{standard1}) 
we may assume that in the $S$-sum $S(i;m_1,...;x_1,...)$ no
$m_j$ is negative and that if some $m_j$ is zero, then the corresponding
$x_j$ is not equal to $1$. 
The sum $S(i,...)$ is then rewritten as
\bq
\label{hoelder}
\lefteqn{
    S(i;m_1,...,m_k;x_1,...,x_k) 
    = S(N;m_1,...,m_k;x_1,...,x_k) 
} & & \nonumber \\
& &
     \mbox{} - S(N;m_2,...,m_k;x_2,...,x_k) \times 
          \left( \sum\limits_{i_1=i+1}^{N} 
                    \frac{x_1^{i_1}}{i_1^{m_1}} \right)
\nonumber \\
& &
     + S(N;m_3,...,m_k;x_3,...,x_k) \times 
          \left( \sum\limits_{i_1=i+1}^{N} \sum\limits_{i_2=i_1+1}^{N} 
                 \frac{x_1^{i_1}}{i_1^{m_1}} \frac{x_2^{i_2}}{i_2^{m_2}} \right) 
\nonumber \\
& &
     \mbox{}  - ... + (-1)^k \left(\sum\limits_{i_1=i+1}^{N} \sum\limits_{i_2=i_1+1}^{N} 
                                      ... \sum\limits_{i_k=i_{k-1}+1}^{N} 
                                    \frac{x_1^{i_1}}{i_1^{m_1}} 
                                    \frac{x_2^{i_2}}{i_2^{m_2}} ...
                                    \frac{x_k^{i_k}}{i_k^{m_k}} \right).
\eq
The proof of eq. (\ref{hoelder}) is not too complicated and consists in repeated
application of the identity
\bq
\label{proofhoelder}
\sum\limits_{i=1}^n \sum\limits_{j=1}^i a_{ij}
& = &
\sum\limits_{i=1}^N \sum\limits_{j=1}^i a_{ij}
- \sum\limits_{i=n+1}^N \sum\limits_{j=1}^N a_{ij}
+ \sum\limits_{i=n+1}^N \sum\limits_{j=i+1}^N a_{ij}.
\eq
Eq. (\ref{hoelder}) holds for any $N$ and in particular we may take $N=\infty$ 
in the end.
Each term is then a product of a $S$-sum at infinity and a sum of a new type.
The $S$-sum at infinity is converted to a $Z$-sum at infinity 
and expressed in terms of multiple
polylogarithms.
We now deal with sums of the form
\bq
\label{Csum}
     \mbox{} - \sum\limits_{i=1}^n 
       \left( \begin{array}{c} n \\ i \\ \end{array} \right)
       \left( -1 \right)^i
       \frac{x_0^i}{i^{m_0}}
       \sum\limits_{i_1=i+1}^N
       \sum\limits_{i_2=i_1+1}^N ...
       \sum\limits_{i_k=i_{k-1}+1}^N
        \frac{x_1^{i_1}}{i_1^{m_1}}
        \frac{x_2^{i_2}}{i_2^{m_2}} ...
        \frac{x_k^{i_k}}{i_k^{m_k}}.
\eq
We introduce raising and lowering operators as follows:
\bq
\left( {\bf x^+} \right)^m \cdot 1 & = & \frac{1}{m!} \ln^m(x), \nonumber \\
{\bf x^+} \cdot f(x) & = & \int\limits_0^x \frac{dx'}{x'} f(x'), \nonumber \\ 
{\bf x^-} \cdot f(x) & = & x \frac{d}{dx} f(x). 
\eq
It is understood that in the second line only functions which are integrable at $x=0$ are
considered. 
We see that
${\bf x^-}$ is the inverse to ${\bf x^+}$, e.g. ${\bf x^- x^+} = \mbox{id}$.
However ${\bf x^+ x^-} = \mbox{id}$ holds only if applied to non-trivial sums.
For the trivial sum we have ${\bf x^+ x^-} Z(n) = 0$. \\
With the help of the raising operators eq. (\ref{Csum}) may be rewritten as
\bq
\label{Csum_raising_0}
\lefteqn{
     \left( {\bf x_k}^+ \right)^{m_k} \left( {\bf x_{k-1}}^+ \right)^{m_{k-1}} ... 
     \left( {\bf x_1}^+ \right)^{m_1} \left( {\bf x_0}^+ \right)^{m_0}
     (-1) \sum\limits_{i=1}^n 
       \left( \begin{array}{c} n \\ i \\ \end{array} \right)
       \left( - x_0 \right)^i
} & & \nonumber \\
& & \times
       \sum\limits_{i_1=i+1}^N
       \sum\limits_{i_2=i_1+1}^N ...
       \sum\limits_{i_k=i_{k-1}+1}^N
        x_1^{i_1}
        x_2^{i_2} ...
        x_k^{i_k}.
\hspace*{5cm}
\eq
It may happen that some $x_i$'s are equal to one. In this case we first calculate
the sum for arbitrary $x_i$'s and take then the limit $x_i \rightarrow 1$.
Some care has to be taken for the double limit $x \rightarrow 1$ and 
$N \rightarrow \infty$. The order is as follows:
First all limits $x \rightarrow 1$ are taken, then the limit
$N \rightarrow \infty$ in eq. (\ref{hoelder}) is performed.\\
\\
The sums in eq. (\ref{Csum_raising_0}) can be performed with the
help of the geometric series
\bq
\label{geometric_series}
\sum\limits_{i=n+1}^N x^i & = & \frac{x}{1-x} x^n - \frac{x}{1-x} x^N.
\eq
It is evident that if we don't have to take the limit $x \rightarrow 1$
we can immediately neglect the second term.
Also in the case $x=1$ the second term can be neglected. It gives rise to terms
of the form
\bq
\label{zero_contr}
\left( {\bf x}^+ \right)^m \frac{x}{1-x} x^N 
 & = &
 \sum\limits_{i=N+1}^\infty \frac{x^i}{i^m}.
\eq
On the r.h.s the limit $x \rightarrow 1$ may safely be performed and
the resulting sum gives a vanishing contribution in the limit $N \rightarrow \infty$.\\
\\
Performing the sums in eq. (\ref{Csum_raising_0}) we therefore only have to consider 
expressions of the form
\bq
\label{Csum_raising}
\lefteqn{
     \left( {\bf x_k}^+ \right)^{m_k} \left( {\bf x_{k-1}}^+ \right)^{m_{k-1}} ... 
     \left( {\bf x_1}^+ \right)^{m_1} \left( {\bf x_0}^+ \right)^{m_0}
     \frac{x_k}{1-x_k} \frac{x_{k-1}x_k}{1-x_{k-1}x_k} ...
     \frac{x_1 ... x_k}{1-x_1 ... x_k}
} & & \nonumber \\
& & \times
     \left[ 1 - \left( 1 - x_0 x_1 ... x_k \right)^n \right].
\hspace*{7cm}
\eq
We then perform succesivly the integrations corresponding to the raising operators.
The basic formulae are:
\bq
\label{raisint}
{\bf x_1^+} \left[ 1 - \left( 1 - x_1 x_2 \right)^{n} \right] 
 & = & \sum\limits_{i=1}^n \frac{1}{i} \left[ 1 - \left( 1- x_1 x_2 \right)^i \right],
\nonumber \\
{\bf x_1^+} \frac{x_1 x_2}{1-x_1 x_2} \left[ 1 - \left( 1 - x_0 x_1 x_2 \right)^n \right] 
 & = & 
   - \left( 1 - x_0 \right)^n \sum\limits_{i=1}^n \frac{1}{i} 
                \left( \frac{1}{1-x_0} \right)^i \left[ 1 - \left( 1 - x_0 x_1 x_2 \right)^i \right] \nonumber \\
& & 
    + \left( 1 - \left( 1 - x_0 \right)^n \right) \sum\limits_{i=1}^N \frac{\left(x_1 x_2 \right)^i}{i} \nonumber \\
& &
+ {\bf x_1^+} \frac{x_1 x_2}{1-x_1 x_2} \left( x_1 x_2 \right)^N 
                                      \left( 1 - \left( 1 - x_0 \right)^n \right),
\nonumber \\
{\bf x_1^+} \frac{x_1 x_2}{1-x_1x_2} \left[ 1 - \left( 1 - x_1 x_2 \right)^{n} \right]
 & = &
 - \frac{1}{n} \left[ 1 - \left( 1 - x_1 x_2 \right)^{n} \right]
 + \sum\limits_{i=1}^N \frac{(x_1 x_2)^i}{i}
\nonumber \\
& &
 + \left( {\bf x_1^+} \right) \frac{x_1 x_2}{1-x_1x_2} \left( x_1 x_2 \right)^N.
\eq
We use the first formula to reduce $m_0$ in eq. (\ref{Csum_raising}) to zero:
\bq
    \sum\limits_{i=1}^{n} \frac{1}{i} 
     \left( x_k^+ \right)^{m_k} ... 
     \left( x_1^+ \right)^{m_1} \left( x_0^+ \right)^{m_0-1}
     \frac{x_k}{1-x_k} \frac{x_{k-1}x_k}{1-x_{k-1}x_k} ...
     \frac{x_1 ... x_k}{1-x_1 ... x_k}
     \left[ 1 - \left( 1 - x_0 x_1 ... x_k \right)^i \right].
\nonumber \\
\eq
In the following we may therefore assume $m_0=0$ in eq. (\ref{Csum_raising}).
If $m_0=0$, $m_1>0$ and $x_0 \neq 1 $ we obtain for eq. (\ref{Csum_raising})
\bq
\lefteqn{
    - (1-x_0)^n 
     \sum\limits_{i=1}^\infty \frac{1}{i} \left( \frac{1}{1-x_0} \right)^i 
     \left( x_k^+ \right)^{m_k} ... 
     \left( \left(x_0 x_1\right)^+ \right)^{m_1-1} 
     \frac{x_k}{1-x_k} \frac{x_{k-1}x_k}{1-x_{k-1}x_k} ...
     \frac{x_2 ... x_k}{1-x_2 ... x_k}
} \nonumber \\
& & \times
     \left[ 1 - \left( 1 - (x_0 x_1) ... x_k \right)^i \right]
\nonumber \\
& &
    + ( 1 - (1-x_0)^n)
      \sum\limits_{i=1}^N \frac{x_1^{i}}{i^{m_1}} 
      \sum\limits_{i_2=i+1}^{N} ... \sum\limits_{i_k = i_{k-1}+1}^{N}
      \frac{x_2^{i_2}}{i_2^{m_2}} ... \frac{x_k^{i_k}}{i_k^{m_k}}.
\hspace*{5cm}
\eq
Here we neglected terms of the form eq. (\ref{zero_contr}).
In the case $m_0=0$, $m_1>0$ and $x_0 = 1 $ we use the third formula of eq. (\ref{raisint}).
Again we may neglect contributions of the form eq. (\ref{zero_contr}).
Doing so we obtain
\bq
\lefteqn{
    \mbox{} - \frac{1}{n} 
     \left( x_k^+ \right)^{m_k} ... 
     \left( x_1^+ \right)^{m_1-1} 
     \frac{x_k}{1-x_k} \frac{x_{k-1}x_k}{1-x_{k-1}x_k} ...
     \frac{x_2 ... x_k}{1-x_2 ... x_k}
     \left[ 1 - \left( 1 - x_1 ... x_k \right)^n \right]
} & & \nonumber \\
& &
    \mbox{} + 
      \sum\limits_{i=1}^N \frac{x_1^{i}}{i^{m_1}} 
      \sum\limits_{i_2=i+1}^{N} ... \sum\limits_{i_k = i_{k-1}+1}^{N}
      \frac{x_2^{i_2}}{i_2^{m_2}} ... \frac{x_k^{i_k}}{i_k^{m_k}}.
\hspace*{5cm}
\eq
It remains to treat the last term to complete the recursion.
The last term introduces a sum of the type
\bq
      \sum\limits_{i_1=n+1}^{N} ... \sum\limits_{i_k = i_{k-1}+1}^{N}
      \frac{x_1^{i_1}}{i_1^{m_1}} ... \frac{x_k^{i_k}}{i_k^{m_k}}.
\eq
Using the inverse formula to eq. (\ref{hoelder})
\bq
\lefteqn{
      \sum\limits_{i_1=n+1}^{N} ... \sum\limits_{i_k = i_{k-1}+1}^{N}
      \frac{x_1^{i_1}}{i_1^{m_1}} ... \frac{x_k^{i_k}}{i_k^{m_k}}
} & & \nonumber \\
& = &
     (-1)^k S(n;m_1,...m_k;x_1,...,x_k) - (-1)^k S(N;m_1,...m_k;x_1,...,x_k)
\nonumber \\
& &
           + (-1)^k S(N;m_2,...m_k;x_2,...,x_k)
                    \sum\limits_{i_1=n+1}^{N} \frac{x_1^{i_1}}{i_1^{m_1}} 
           - ...
\nonumber \\
& &
           + (-1)^k S(N;m_k;x_k)
                    \sum\limits_{i_1=n+1}^{N} ... \sum\limits_{i_{k-1} = i_{k-2}+1}^{N}
                    \frac{x_1^{i_1}}{i_1^{m_1}} ... \frac{x_{k-1}^{i_{k-1}}}{i_{k-1}^{m_{k-1}}}
\eq
this sum is easily related to $S$-sums with upper summation limit $n$ 
and (after taking the limit $N \rightarrow \infty$) to multiple polylogarithms.\\
\\
If $m_0=0$, $m_1=0$ and $x_1 \neq 1$ we rewrite eq. (\ref{Csum_raising}) as
\bq
\lefteqn{
    -\frac{1}{1-x_1} 
     \left( x_k^+ \right)^{m_k} ... 
     \left( x_3^+ \right)^{m_3} 
     \left( \left(x_2 x_1\right)^+ \right)^{m_2} 
     \frac{x_k}{1-x_k} ...
     \frac{x_3 ... x_k}{1-x_3 ... x_k}
     \frac{(x_1 x_2) x_3 ... x_k}{1-(x_1 x_2) x_3 ... x_k}
} & & \nonumber \\
& & \times
     \left[ 1 - \left( 1 - x_0 (x_1 x_2) x_3 ... x_k \right)^i \right]
\nonumber \\
& &
    \mbox{} + \frac{x_1}{1-x_1} 
     \left( x_k^+ \right)^{m_k} ... 
     \left( x_2^+ \right)^{m_2} 
     \frac{x_k}{1-x_k} ...
     \frac{x_2 x_3 ... x_k}{1-x_2 x_3 ... x_k}
     \left[ 1 - \left( 1 - (x_0 x_1) x_2 x_3 ... x_k \right)^i \right].
\nonumber \\
\eq
The case $m_0=0$, $m_1=0$ and $x_1=1$ has to be excluded.
However, with an appropriate choice of the standard form for $S$-sums (c.f. eq. \ref{standard1}) 
this case never occurs.

\subsection{Algorithm D}
\label{subsubsec:type_D}

Here we consider sums of the form
\bq
\lefteqn{
     \mbox{} - \sum\limits_{i=1}^{n-1}
       \left( \begin{array}{c} n \\ i \\ \end{array} \right)
       \left( -1 \right)^i
       \frac{x^i}{(i+c)^m} 
       \frac{\Gamma(i+a_1+b_1\varepsilon)}{\Gamma(i+c_1+d_1\varepsilon)}
       ...
       \frac{\Gamma(i+a_k+b_k\varepsilon)}{\Gamma(i+c_k+d_k\varepsilon)}
} & &
\nonumber \\
& &
       \times
       S(i+o,m_1,...,m_l,x_1,...,x_l)
\nonumber \\
& & \times
       \frac{y^{n-i}}{(n-i+c')^{m'}} 
       \frac{\Gamma(n-i+a_1'+b_1'\varepsilon)}{\Gamma(n-i+c_1'+d_1'\varepsilon)}
       ...
       \frac{\Gamma(n-i+a_{k'}'+b_{k'}'\varepsilon)}{\Gamma(n-i+c_{k'}'+d_{k'}'\varepsilon)}
\nonumber \\
& & \times
       S(n-i+o',m_1',...,m_{l'}',x_1',...,x_{l'}').
\eq
Here, all $a_j$, $a_j'$, $c_j$  and $c_j'$ are integers, $c$, $c'$, 
are nonnegative integers and $o $,
$o' $ are integers.
Note that the upper summation limit is $(n-1)$.
The upper summation limit should not be infinity.
As in the case of sums of type C, we cannot relate these sums to
$Z$-sums with upper summation limit $(n-1)$ alone, but
we can reduce them to $Z$-sums with upper summation limit $(n-1)$ and
multiple polylogarithms (which are $Z$-sums to infinity).
\\
\\
After the expansion of the Gamma functions and the synchronization of the subsums
we have sums of the form
\bq
    \mbox{} - \sum\limits_{i=1}^{n-1} 
       \left( \begin{array}{c} n \\ i \\ \end{array} \right)
       \left( -1 \right)^i
       \frac{x^i}{(i+c)^m} S(i,m_1,...) \frac{y^{n-i}}{(n-i+c')^{m'}} S(n-i,m_1',....).
\eq
Partial fractioning leads to
\bq
\label{type_D_letter_offset}
    \mbox{} - \sum\limits_{i=1}^{n-1} 
       \left( \begin{array}{c} n \\ i \\ \end{array} \right)
       \left( -1 \right)^i
       \frac{x^i}{(i+c)^m} S(i,m_1,...) S(n-i,m_1',....)
\eq
with $c \ge 0$. 
In order to 
reduce the offset $c$ to zero one rewrites eq. (\ref{type_D_letter_offset}) as
\bq
    \left(-\frac{1}{x}\right) \frac{1}{n+1}
     (-1) \sum\limits_{i=1}^{n+1-1} 
       \left( \begin{array}{c} n+1 \\ i \\ \end{array} \right)
       \left( -1 \right)^i
      \frac{x^i}{(i+c-1)^m} i S(i-1,...) S(n+1-i,m_1',....).
\nonumber \\
\eq
We arrive at sums of the form
\bq
\label{type_D_one_letter}
    \mbox{} - \sum\limits_{i=1}^{n-1} 
       \left( \begin{array}{c} n \\ i \\ \end{array} \right)
       \left( -1 \right)^i
       \frac{x^i}{i^m} S(i,m_1,...) S(n-i,m_1',....).
\eq
If the depth of $S(n-i,m_1',....)$ is zero, we have a sum of type C:
\bq
\lefteqn{
    \mbox{} - \sum\limits_{i=1}^{n-1} 
       \left( \begin{array}{c} n \\ i \\ \end{array} \right)
       \left( -1 \right)^i
       \frac{x^i}{i^m} S(i,m_1,...)
} & & \nonumber \\
& = &
    \mbox{} - \sum\limits_{i=1}^{n} 
       \left( \begin{array}{c} n \\ i \\ \end{array} \right)
       \left( -1 \right)^i
       \frac{x^i}{i^m} S(i,m_1,...)
 + \frac{(-x)^n}{n^m} S(n,m_1,...).
\eq
Otherwise, we first reduce $m$ to zero.
For $m>0$ we rewrite eq. (\ref{type_D_one_letter}) as
\bq
\lefteqn{
    \sum\limits_{j=1}^n \frac{1}{j} 
     \left[ (-1) \sum\limits_{i=1}^{j-1}
       \left( \begin{array}{c} j \\ i \\ \end{array} \right)
       \left( -1 \right)^i
       \frac{x^i}{i^{m-1}} S(i,m_1,...) S(j-i,m_1',....)
     \right]
} & & \nonumber \\
& &
     \mbox{} +
    \sum\limits_{j=1}^n \frac{1}{j} 
     \left[ (-1) \sum\limits_{i=1}^{j-1}
       \left( \begin{array}{c} j \\ i \\ \end{array} \right)
       \left( -1 \right)^i
       \frac{x^i}{i^{m}} S(i,m_1,...) \frac{{x_1'}^{j-i}}{(j-i)^{m_1'-1}} S(j-i,m_2',....)
     \right].
\eq
For $m<0$ we rewrite eq. (\ref{type_D_one_letter}) as
\bq
\lefteqn{
    \mbox{} - n x (-1) \sum\limits_{i=1}^{n-2} 
       \left( \begin{array}{c} n-1 \\ i \\ \end{array} \right)
       \left( -1 \right)^i
       \frac{x^i}{(i+1)^{m+1}} S(i+1,m_1,...) S(n-1-i,m_1',....)
} & & \nonumber \\
& & + n x S(1,m_1,...) S(n-1,m_1',...).
\hspace*{6cm}
\eq
Having reduced $m$ to zero we arrive at sums of the form
\bq
    \mbox{} - \sum\limits_{i=1}^{n-1} 
       \left( \begin{array}{c} n \\ i \\ \end{array} \right)
       \left( -1 \right)^i
       x^i S(i,m_1,...) S(n-i,m_1',....).
\eq
For $x \neq 1 $ we obtain after some algebra
\bq
\label{new_eq71}
\lefteqn{
    \mbox{} - \sum\limits_{i=1}^{n-1} 
       \left( \begin{array}{c} n \\ i \\ \end{array} \right)
       \left( -1 \right)^i
       x^i S(i,m_1,...) S(n-i,m_1',....) } & & 
\nonumber \\
    & = & (1-x)^n \sum\limits_{j=1}^n \frac{1}{j} \left( \frac{1}{1-x} \right)^j 
       (-1) \sum\limits_{i=1}^{j-1} 
       \left( \begin{array}{c} j \\ i \\ \end{array} \right)
       \left( -1 \right)^i
       \frac{(x x_1)^i}{i^{m_1-1}} S(i,m_2,...) S(j-i,m_1',....)
\nonumber \\
& & 
    + (1-x)^n \sum\limits_{j=1}^n \frac{1}{j} \left( \frac{1}{1-x} \right)^j
       (-1) \sum\limits_{i=1}^{j-1} 
       \left( \begin{array}{c} j \\ i \\ \end{array} \right)
       \left( -1 \right)^i
       x^i S(i,m_1,...) \frac{{x_1'}^{j-i}}{(n-i)^{m_1'-1}} S(j-i,m_2',....),
\nonumber \\
\eq
where the original problem is reduced to one of the same type but
with lower weight.
In the case $x=1$ the r.h.s of eq. (\ref{new_eq71}) reduces to
\bq
\lefteqn{
    \frac{1}{n} 
       (-1) \sum\limits_{i=1}^{n-1} 
       \left( \begin{array}{c} n \\ i \\ \end{array} \right)
       \left( -1 \right)^i
       \frac{x_1^i}{i^{m_1-1}} S(i,m_2,...) S(n-i,m_1',....)
} & & \nonumber \\
& &
    + \frac{1}{n} 
       (-1) \sum\limits_{i=1}^{n-1} 
       \left( \begin{array}{c} n \\ i \\ \end{array} \right)
       \left( -1 \right)^i
       S(i,m_1,...) \frac{{x_1'}^{n-i}}{(n-i)^{m_1'-1}} S(n-i,m_2',....).
\eq
Again, the original problem is reduced to one of the same type but
with lower weight.
The above algorithm yields thus a recursion to treat sums of type D.

\section{Applications}
\label{sec:applications}

The algorithms given in this paper can be used for the expansion of
higher transcendental functions around integer 
values of their indices, where the expansion
parameter occurs in the Pochhammer symbols.
In this section we give a few examples. 
Additionally, we illustrate the applicability of the algorithms for nested sums 
to the calculation of loop integrals, in particular to integrals with several 
scales. As an example, we discuss the C-topology and show that the integral 
can be written as a nested sum of the type previously discussed.

\subsection{Generalized hypergeometric functions}

The generalized hypergeometric functions are defined by \cite{Erdelyi,Slater}
\bq
_{J+1}F_J(a_1,...,a_{J+1};b_1,...,b_J;x) & = & \sum\limits_{n=0}^{\infty} 
\frac{(a_1)_n ... (a_{J+1})_n}{(b_1)_n ... (b_J)_n}
 \frac{x^{n}}{n!},
\eq
where $(a)_n = \Gamma(n+a)/ \Gamma(a)$ denotes a Pochhammer symbol. 
These functions can be rewritten as
\bq
\lefteqn{
_{J+1}F_J(a_1,...,a_{J+1};b_1,...,b_J;x) 
} & & \nonumber \\
 & = & 
 1 + \frac{\Gamma(b_1) ... \Gamma(b_J)}{\Gamma(a_1) ... \Gamma(a_{J+1})} 
 \sum\limits_{i=1}^{\infty} x^i 
   \frac{\Gamma(i+a_1)}{\Gamma(i+b_1)} ...
   \frac{\Gamma(i+a_J)}{\Gamma(i+b_J)} 
   \frac{\Gamma(i+a_{J+1})}{\Gamma(i+1)} 
\eq
and fall therefore into the category of transcendental sums of type A.
We give a few examples obtained 
using the algorithms given in sec. \ref{subsubsec:type_A}:
\bq
{}_2F_{1}( a \eps, b \eps; 1 - c \eps; x) & = & 
  1 + a b \;\mbox{Li}_2(x) \eps^2
  + a b \left( c \;\mbox{Li}_3(x) + ( a+b+c ) \;\mbox{S}_{1,2}(x) \right) \eps^3
  + O(\eps^4), \nonumber \\
{}_2F_{1}( 1, -\eps; 1 - \eps; x) & = & 
  1 + \ln(1-x) \eps - \;\mbox{Li}_2(x) \eps^2
  - \;\mbox{Li}_3(x) \eps^3
  - \;\mbox{Li}_4(x) \eps^4
  - \;\mbox{Li}_5(x) \eps^5
\nonumber \\
& &
  - \;\mbox{Li}_6(x) \eps^6
  - \;\mbox{Li}_7(x) \eps^7 + O(\eps^8),
\eq
\bq
\lefteqn{
{}_3F_{2}( -2\eps, -2\eps, 1-\eps; 1-2\eps, 1-2\eps; x) =  
  1 + 4 \;\mbox{Li}_2(x) \eps^2 
  + \left( 12 \;\mbox{Li}_3(x) - 4 S_{1,2}(x) \right) \eps^3
} & & \nonumber \\
& & 
  + \left( 32 \;\mbox{Li}_4(x) + 4 S_{1,3}(x) -12 S_{2,2}(x) \right) \eps^4
  + \left( 80 \;\mbox{Li}_5(x) - 4 S_{1,4}(x) +12 S_{2,3}(x) - 32 S_{3,2}(x) \right) \eps^5
\nonumber \\ & &
  + O(\eps^6),
\eq
which all agree with known results in the literature 
\cite{Kosower:1999rx,Anastasiou:1999cx}.

\subsection{Appell functions}

The first Appell function is defined by \cite{Appell,Slater}
\bq
F_1(a,b_1,b_2;c;x_1,x_2)  
& = & \sum\limits_{m_1=0}^\infty \sum\limits_{m_2=0}^\infty 
\frac{(a)_{m_1+m_2} (b_1)_{m_1} (b_2)_{m_2}}
{(c)_{m_1 + m_2}}
\frac{x_1^{m_1}}{m_1!} \frac{x_2^{m_2}}{m_2!}.
\eq
It can be rewritten as
\bq
\lefteqn{
F_1(a,b_1,b_2;c;x_1,x_2)  
} & & \nonumber \\
& = & 1
+ \frac{\Gamma(c)}{\Gamma(a) \Gamma(b_1)}
    \sum\limits_{i=1}^\infty x_1^i \frac{\Gamma(i+a) \Gamma(i+b_1)}{\Gamma(i+c) \Gamma(i+1)}
+ \frac{\Gamma(c)}{\Gamma(a) \Gamma(b_2)}
    \sum\limits_{i=1}^\infty x_2^i \frac{\Gamma(i+a) \Gamma(i+b_2)}{\Gamma(i+c) \Gamma(i+1)}
\nonumber \\
& &
+ \frac{\Gamma(c)}{\Gamma(a) \Gamma(b_1) \Gamma(b_2)}
    \sum\limits_{n=1}^\infty \frac{\Gamma(n+a)}{\Gamma(n+c)}
    \sum\limits_{i=1}^{n-1} x_1^i \frac{\Gamma(i+b_1)}{\Gamma(i+1)}
                            x_2^{n-i} \frac{\Gamma(n-i+b_2)}{\Gamma(n-i+1)}.
\eq
The inner sum of the last term is of type B. The first Appell function can
therefore be expanded with the help of the algorithms A and B.\\
\\
The second Appell function is defined by \cite{Appell,Slater}
\bq
F_2(a,b_1,b_2;c_1,c_2;x_1,x_2)  
& = & \sum\limits_{m_1=0}^\infty \sum\limits_{m_2=0}^\infty 
\frac{(a)_{m_1+m_2} (b_1)_{m_1} (b_2)_{m_2}}
{(c_1)_{m_1} (c_2)_{m_2}}
\frac{x_1^{m_1}}{m_1!} \frac{x_2^{m_2}}{m_2!}.
\eq
It can be rewritten as
\bq
\lefteqn{
F_2(a,b_1,b_2;c_1,c_2;x_1,x_2)  
} & & \nonumber \\
& = & 1
+ \frac{\Gamma(c_1)}{\Gamma(a) \Gamma(b_1)}
    \sum\limits_{i=1}^\infty x_1^i \frac{\Gamma(i+a) \Gamma(i+b_1)}{\Gamma(i+c_1) \Gamma(i+1)}
+ \frac{\Gamma(c_2)}{\Gamma(a) \Gamma(b_2)}
    \sum\limits_{i=1}^\infty x_2^i \frac{\Gamma(i+a) \Gamma(i+b_2)}{\Gamma(i+c_2) \Gamma(i+1)}
\nonumber \\
& &
- \frac{\Gamma(c_1) \Gamma(c_2)}{\Gamma(a) \Gamma(b_1) \Gamma(b_2)}
    \sum\limits_{n=1}^\infty \frac{\Gamma(n+a)}{\Gamma(n+1)}
\nonumber \\
& &
    \times (-1) 
    \sum\limits_{i=1}^{n-1} 
       \left( \begin{array}{c} n \\ i \\ \end{array} \right)
       \left( -1 \right)^i
                            \left(-x_1 \right)^i \frac{\Gamma(i+b_1)}{\Gamma(i+c_1)}
                            x_2^{n-i} \frac{\Gamma(n-i+b_2)}{\Gamma(n-i+c_2)}.
\eq
The inner sum of the last term is of type D. The second Appell function can
therefore be expanded with the help of the algorithms A to D.
As an example we give
\bq
\label{expansionF2}
 &&
 F_2(1,1,\eps;1+\eps,1-\eps;x,y) =  
  \frac{1}{1-x}
  + \frac{1}{1-x} \left( 2 \ln(1-x) - \ln(1-x-y) \right) \eps 
 \nonumber \\
&&
 + \frac{1}{1-x} \left[ 2 \;\mbox{Li}_2(x) + 2 \;\mbox{Li}_2(y) - \;\mbox{Li}_2(x+y)
                        + 4 S_{0,2}(x) + S_{0,2}(x+y)
                        + \;\mbox{Li}_{1,1}\left( \frac{y}{x+y}, x+y \right)
\right. \nonumber \\ & & \left.
                        -2 \;\mbox{Li}_{1,1}\left( \frac{x}{x+y}, x+y \right)
                        -2 \;\mbox{Li}_{1,1}\left( \frac{x+y}{x}, x \right)
                 \right] \eps^2
 \nonumber \\
 &&
 + \frac{1}{1-x} \Bigg[
  \mbox{Li}_3(x+y)
  - 2 \;  \mbox{Li}_3(x)
  + S_{0,3}(x+y)
  - 8 S_{0,3}(x)
  - S_{1,2}(x+y)
  - 4 S_{1,2}(x)
  + 4 S_{1,2}(y)\nonumber\\
  &&
  - H_{1,2}(x+y)
  - 4 H_{1,2}(x)
  + 2 \; \mbox{Li}_{1,2}\left( \frac{x}{x+y}, x+y \right)
  -  \; \mbox{Li}_{1,2}\left( \frac{y}{x+y}, x+y \right)\nonumber\\
  &&
  + 2 \; \mbox{Li}_{1,2}\left( \frac{x+y}{x}, x \right)
  + 2 \; \mbox{Li}_{2,1}\left( \frac{x}{x+y}, x+y \right)
  + 3 \; \mbox{Li}_{2,1}\left( \frac{y}{x+y}, x+y \right)
  + 2 \; \mbox{Li}_{2,1}\left( \frac{x+y}{x}, x \right)\nonumber\\
  &&
  - 4 \; \mbox{Li}_{2,1}\left( \frac{y}{x}, x \right)
  + 4 \; \mbox{Li}_{1,1,1}\left( \frac{x}{x+y}, \frac{x+y}{x}, x \right)
  + 4 \; \mbox{Li}_{1,1,1}\left( \frac{x+y}{x}, 1, x \right)\nonumber\\
  &&
  + 4 \; \mbox{Li}_{1,1,1}\left( 1, \frac{x}{x+y}, x+y \right)
  + 2 \; \mbox{Li}_{1,1,1}\left( 1, \frac{y}{x+y}, x+y \right)
  - 2 \; \mbox{Li}_{1,1,1}\left( \frac{y}{x+y}, \frac{x+y}{x}, x \right)
  \nonumber\\
  &&
  +   \; \mbox{Li}_{1,1,1}\left( \frac{y}{x+y}, 1, x+y \right)
  - 2 \; \mbox{Li}_{1,1,1}\left( \frac{x}{x+y}, 1, x+y \right)
  - 2 \; \mbox{Li}_{1,1,1}\left( 1, \frac{x+y}{x}, x \right)\nonumber\\
  &&
  - 2 \; \mbox{Li}_{1,1,1}\left( \frac{x+y}{x}, \frac{x}{x+y}, x+y \right)
 \Bigg] \eps^3
 + O(\eps^4).
\eq
After taking into account a typo in eq. (A.47) of ref. \cite{Anastasiou:1999ui}
this result agrees up to order $\eps$ with the one obtained along 
the lines of ref. \cite{Anastasiou:1999ui}.
Multiple polylogarithms of low weight can be expressed as products of classical polylogarithms
and the result of the expansion in eq. (\ref{expansionF2}) can be 
simplified \footnote{We would like to thank the referee for pointing this out.}.
However, we present here the output of our algorithm. Our algorithm gives the result
as a linear combination of polylogarithms and does not generate products of polylogarithms.

\subsection{Kamp\'e de F\'eriet function}

The Kamp\'e de F\'eriet function is defined by \cite{Appell}
\bq
S_1(a_1,a_2,b_1;c,c_1;x_1,x_2)  
& = & \sum\limits_{m_1=0}^\infty \sum\limits_{m_2=0}^\infty 
\frac{(a_1)_{m_1+m_2} (a_2)_{m_1+m_2} (b_1)_{m_1} }
{(c)_{m_1 + m_2} (c_1)_{m_1} }
\frac{x_1^{m_1}}{m_1!} \frac{x_2^{m_2}}{m_2!}.
\eq
It can be rewritten as
\bq
\lefteqn{
S_1(a_1,a_2,b_1;c,c_1;x_1,x_2)  
} & & \nonumber \\
& = & 1
+ \frac{\Gamma(c) \Gamma(c_1)}{\Gamma(a_1) \Gamma(a_2) \Gamma(b_1)}
    \sum\limits_{i=1}^\infty x_1^i \frac{\Gamma(i+a_1) \Gamma(i+a_2) \Gamma(i+b_1)}{\Gamma(i+c) \Gamma(i+c_1) \Gamma(i+1)}
\nonumber \\
& &
+ \frac{\Gamma(c)}{\Gamma(a_1) \Gamma(a_2)}
    \sum\limits_{i=1}^\infty x_2^i \frac{\Gamma(i+a_1) \Gamma(i+a_2)}{\Gamma(i+c) \Gamma(i+1)}
\nonumber \\
& &
- \frac{\Gamma(c) \Gamma(c_1) }{\Gamma(a_1) \Gamma(a_2) \Gamma(b_1)}
    \sum\limits_{n=1}^\infty x_2^n \frac{\Gamma(n+a_1) \Gamma(n+a_2)}{\Gamma(n+c) \Gamma(n+1)}
\nonumber \\
& &
    \times (-1) 
    \sum\limits_{i=1}^{n-1} 
       \left( \begin{array}{c} n \\ i \\ \end{array} \right)
       \left( -1 \right)^i
                            \left(- \frac{x_1}{x_2} \right)^i \frac{\Gamma(i+b_1)}{\Gamma(i+c_1)}.
\eq
The inner sum of the last term is of type C. The Kamp\'e de F\'eriet function can
therefore be expanded with the help of the algorithms A and C.

\subsection{The C-topology}
\label{subsec:Ctopo}

\begin{figure}
\begin{center}
\begin{eqnarray*}
\begin{picture}(100,50)(20,45)
\Vertex(50,20){2}
\Vertex(50,80){2}
\Vertex(20,50){2}
\Vertex(80,50){2}
\ArrowLine(20,50)(50,80)
\ArrowLine(50,20)(20,50)
\ArrowLine(50,80)(80,50)
\ArrowLine(80,50)(50,20)
\ArrowLine(50,20)(50,80)
\ArrowLine(50,80)(70,80)
\ArrowLine(50,20)(70,20)
\ArrowLine(80,50)(100,50)
\Text(75,80)[l]{$p_1$}
\Text(105,50)[l]{$p_2$}
\Text(75,20)[l]{$p_3$}
\Line(10,50)(20,50)
\Line(13,57)(20,50)
\Line(13,43)(20,50)
\Text(32,65)[br]{$k_1$}
\Text(72,65)[l]{$l_2$}
\Text(72,35)[l]{$l_3$}
\Text(35,35)[tr]{$k_4$}
\Text(53,50)[l]{$l_5$}
\end{picture}
& = & 
c \int\limits_0^1 da \; a^{\nu_2-1} \bar{a}^{\nu_3-1}
\begin{picture}(100,50)(0,45)
\Vertex(30,50){2}
\Vertex(70,70){2}
\Vertex(70,30){2}
\ArrowLine(30,50)(70,70)
\ArrowLine(70,70)(70,30)
\ArrowLine(70,30)(30,50)
\Line(20,50)(30,50)
\Line(23,57)(30,50)
\Line(23,43)(30,50)
\Line(70,70)(80,77)
\Line(70,70)(80,63)
\Text(85,70)[l]{$p_1+\bar{a}p_2$}
\Line(70,30)(80,37)
\Line(70,30)(80,23)
\Text(85,30)[l]{$p_3+ap_2$}
\end{picture}
\end{eqnarray*}
\caption{\label{I3m} The C-topology reduces to a triangle with three external masses and
an additional integration over the Feynman parameter $a$.}
\end{center}
\end{figure}
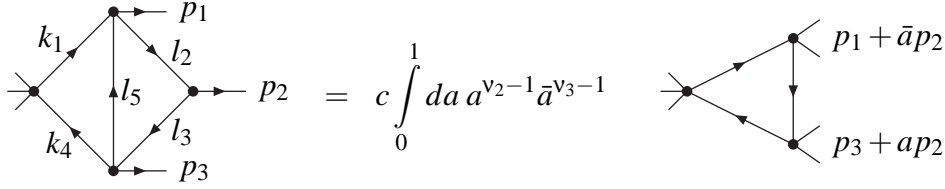

Here we study the C-topology with one massive external leg and arbitrary powers of the 
propagators and arbitray dimensions, which can be solved 
using the algorithms given in this paper.
The importance of this result lies in the fact that one avoids having to solve
a system of equations obtained from partial integration or
Lorentz invariance.
Solving such a system becomes quite difficult if one external leg is massive.
The result obtained here is thus 
a useful ingredient for the calculation of the two-loop
amplitudes with one massive external leg, such as $e^+ e^- \rightarrow 3 \;\mbox{jets}$.
\\
The second C-topology, where the massive leg is attached to the other corner, as well as
all simpler topologies, can be obtained along the same lines.
The two-loop C-topology with one massive external leg is defined by
\bq
I & = & \int \frac{d^Dk_1}{i \pi^{D/2}} \int \frac{d^Dl_5}{i \pi^{D/2}}
 \frac{1}{\left(-k_1^2\right)^{\nu_1}}
 \frac{1}{\left(-l_2^2\right)^{\nu_2}}
 \frac{1}{\left(-l_3^2\right)^{\nu_3}}
 \frac{1}{\left(-k_4^2\right)^{\nu_4}}
 \frac{1}{\left(-l_5^2\right)^{\nu_5}}
\eq
with
\bq
l_2 & = & k_1 + l_5 - p_1, \nonumber \\
l_3 & = & l_2 - p_2, \nonumber \\
k_4 & = & k_1 - p_{123}.
\eq
Fig. \ref{I3m} shows the corresponding Feynman diagram. 
We first perform the $l_5$-integration. Combining with Feynman parameters first
$l_2^2$ and $l_3^2$ and then the resulting propagator with $l_5^2$ we obtain:
\bq
I & = & 
 \frac{\Gamma(\nu_{235}-m+\eps)}{\Gamma(\nu_2) \Gamma(\nu_3) \Gamma(\nu_5)}
 \frac{\Gamma(-\nu_{23}+m-\eps) \Gamma(-\nu_5+m-\eps)}{\Gamma(-\nu_{235}+2m-2\eps)}
 \int\limits_0^1 da \; a^{\nu_2-1} \bar{a}^{\nu_3-1} \nonumber \\
 & & \times \int \frac{d^Dk_1}{i \pi^{D/2}} 
 \frac{1}{\left(-k_1^2\right)^{\nu_1}}
 \frac{1}{\left(-\left(k_1-p_1-\bar{a}p_2\right)^2\right)^{\nu_{235}-m+\eps}}
 \frac{1}{\left(-k_4^2\right)^{\nu_4}}.
\eq
As a short hand notation we used $D=2m-2\eps$, $\bar{a}=1-a$ and $\nu_{235}=\nu_2+\nu_3+\nu_5$.
The second line is a one-loop triangle with three external masses.
The solution for this one-loop integral
with arbitrary powers of the propagators and arbitrary dimensions
is known \cite{Boos:1991rg}.
We use the solution given in \cite{Anastasiou:1999ui} and perform the remaining integration.
We obtain
\bq
\label{result_Ctopo}
I & = & 
 \frac{\Gamma(2m-2\eps-\nu_{1235})\Gamma(1+\nu_{1235}-2m+2\eps)
       \Gamma(2m-2\eps-\nu_{2345})\Gamma(1+\nu_{2345}-2m+2\eps)
      }{\Gamma(\nu_1) \Gamma(\nu_2) \Gamma(\nu_3) \Gamma(\nu_4) \Gamma(\nu_5)
        \Gamma(3m-3\eps-\nu_{12345})} \nonumber \\
& & 
 \times
 \frac{ \Gamma(m-\eps-\nu_5) \Gamma(m-\eps-\nu_{23})}{\Gamma(2m-2\eps-\nu_{235})}
 \left( -s_{123} \right)^{2m-2\eps-\nu_{12345}}
 \sum\limits_{i_1=0}^\infty
 \sum\limits_{i_2=0}^\infty
 \frac{x_1^{i_1}}{i_1!}
 \frac{x_2^{i_2}}{i_2!} \nonumber \\
& & \times \left[
   \frac{\Gamma(i_1+\nu_3) \Gamma(i_2+\nu_2) \Gamma(i_1+i_2-2m+2\eps+\nu_{12345})
                                             \Gamma(i_1+i_2-m+\eps+\nu_{235})
        }{\Gamma(i_1+1-2m+2\eps+\nu_{1235}) \Gamma(i_2+1-2m+2\eps+\nu_{2345})
          \Gamma(i_1+i_2+\nu_{23})}
 \right. \nonumber \\
 & &
 - x_1^{2m-2\eps-\nu_{1235}} 
  \nonumber \\
 & & \times
   \frac{\Gamma(i_1+2m-2\eps-\nu_{125}) \Gamma(i_2+\nu_2) \Gamma(i_1+i_2+\nu_4)
                                             \Gamma(i_1+i_2+m-\eps-\nu_1)
        }{\Gamma(i_1+1+2m-2\eps-\nu_{1235}) \Gamma(i_2+1-2m+2\eps+\nu_{2345})
          \Gamma(i_1+i_2+2m-2\eps-\nu_{15})}
 \nonumber \\
 & &
 - x_2^{2m-2\eps-\nu_{2345}}
 \nonumber \\
 & & \times
   \frac{\Gamma(i_1+\nu_3) \Gamma(i_2+2m-2\eps-\nu_{345}) \Gamma(i_1+i_2+\nu_1)
                                             \Gamma(i_1+i_2+m-\eps-\nu_4)
        }{\Gamma(i_1+1-2m+2\eps+\nu_{1235}) \Gamma(i_2+1+2m-2\eps-\nu_{2345})
          \Gamma(i_1+i_2+2m-2\eps-\nu_{45})}
 \nonumber \\
 & & 
 + x_1^{2m-2\eps-\nu_{1235}} x_2^{2m-2\eps-\nu_{2345}}
   \frac{\Gamma(i_1+2m-2\eps-\nu_{125}) \Gamma(i_2+2m-2\eps-\nu_{345}) 
        }{\Gamma(i_1+1+2m-2\eps-\nu_{1235}) \Gamma(i_2+1+2m-2\eps-\nu_{2345})
          } 
 \nonumber \\
 & & \left. \times
 \frac{
                                          \Gamma(i_1+i_2+2m-2\eps-\nu_{235})
                                          \Gamma(i_1+i_2+3m-3\eps-\nu_{12345})
      }{\Gamma(i_1+i_2+4m-4\eps-\nu_{12345}-\nu_5)}
 \right],
\eq
where we set $x_1=(-s_{12})/(-s_{123})$ and $x_2=(-s_{23})/(-s_{123})$.
Changing the summation indices as in the case of the second Appell function
yields a sum of type D.
For specific (integer) values of $\nu_i$ and $m$ this expression can be expanded
in $\eps$ with the algorithm given in sec. \ref{sec:algo}.
This is a new and useful result.
Up to now this integral has only been known for $m=2$ and the sets
$(1,1,1,1,1)$ and $(1,1,1,1,2)$ for $(\nu_1,\nu_2,\nu_3,\nu_4,\nu_5)$.
We have verified that our result agrees in these two specific cases with 
the ones given by Gehrmann and Remiddi \cite{Gehrmann:2000zt}.

\section{Conclusions}

In this paper we studied some algebraic properties of nested
sums. Based on these properties we developed a number of algorithms
which can be used to expand a certain class of
mathematical functions. 
All presented algorithms are suitable for the implementation on a computer.
These algorithms allow the evaluation of 
integrals occuring in high-energy physics.
As an application we have shown how the two-loop C-topology
can be evaluated for arbitrary powers of the propagators
and arbitrary dimensions.
Furthermore we have shown that the nested sums satisfy a
Hopf algebra and established the connection with the Hopf 
algebra of Kreimer.

\subsection*{Acknowledgements}
\label{sec:acknowledgements}

We would like to thank Jos Vermaseren for useful discussions and the referee for his
detailed remarks.

\begin{appendix}

\section{The Hopf algebra of $Z$-sums}
\label{sec:hopf}

In this section we show that the $Z$-sums form a Hopf algebra \cite{Hopf,Milnor}. 
It is sufficient
to demonstrate that the $Z$-sums form a quasi-shuffle algebra. A
general theorem \cite{Hoffman}
guarantees then, that they also form a Hopf algebra.
We also discuss the connection with the Hopf algebra of Kreimer \cite{Kreimer:1998dp}.\\
\\
Before we start, we have to introduce some notation.
We call a pair $(m_j;x_j)$ a letter and the set of all letters the alphabet $A$. We further call $m_j$ the degree of the letter $(m_j,x_j)$.
On the alphabet $A$ we define a multiplication 
\bq
\label{alphabet}
\cdot & : & A \times A \rightarrow A, \nonumber \\
& & (m_1,x_1) \cdot (m_2,x_2) = (m_1+m_2, x_1 x_2),
\eq
e.g. the $x_j$'s are multiplied and the degrees are added.
As a short-hand notation we will in the following denote a letter just 
by $X_j=(m_j;x_j)$. 
A word is an ordered sequence of letters, e.g.
\bq
W & = & X_1, X_2, ..., X_k.
\eq
We denote the word of length zero by $e$.
The sums defined in (\ref{definition}) are therefore completely
specified by the upper summation limit $n$ 
and a word $W$.
In particular for any positive $n$ the sum corresponding to the empty word $e$ equals 1.
A quasi-shuffle algebra ${\cal A}$ on the vectorspace of words 
is defined by \cite{Hoffman} 
\bq 
\label{algebra}
e \circ W & = & W \circ e = W, \nonumber \\
(X_1,W_1) \circ (X_2,W_2) & = & X_1,(W_1 \circ (X_2,W_2)) + X_2,((X_1,W_1) \circ W_2) \nonumber \\
& & + (X_1 \cdot X_2),(W_1 \circ W_2).
\eq
Note that ``$\cdot$'' denotes multiplication of letters as defined 
in eq. (\ref{alphabet}),
whereas ``$\circ$'' denotes the product in the algebra ${\cal A}$, recursively
defined in eq. (\ref{algebra}).
We observe that the formula for the multiplication of $Z$-sums eq. (\ref{Zmultiplication})
is identical to eq. (\ref{algebra}).
The $Z$-sums form therefore a quasi-shuffle algebra.\\
\\
We now discuss the connection with the Hopf algebra of Kreimer \cite{Kreimer:1998dp}.
Kreimer showed that the process of renormalization of UV-divergences occuring
in quantum field theories
can be formulated in terms of a Hopf algebra structure.
We first recall the properties of an algebra and a coalgebra:
An algebra has a unit and a multiplication, whereas a coalgebra has a counit and a
comultiplication.
A Hopf algebra is an algebra and a coalgebra at the same time, such that the two structures
are compatible with each other. In addition there is an antipode.
We show that the coalgebra structure of $Z$-sums is identical to the coalgebra
structure of the Hopf algebra of Kreimer.
To this aim we introduce the explicit definitions of the
counit, the coproduct and the antipode.
It is convenient to phrase the coalgebra structure in terms of rooted trees.
A $Z$-sums can be represented as rooted trees without any sidebranchings. 
As a concrete example we write down the pictorial representation of a sum 
of depth three :
\\
\vspace*{-15mm}
\bq
Z(n;m_1,m_2,m_3;x_1,x_2,x_3) 
& = &
\sum\limits_{i_1=1}^n
\sum\limits_{i_2=1}^{i_1-1}
\sum\limits_{i_3=1}^{i_2-1}
 \frac{x_1^{i_1}}{{i_1}^{m_1}}
 \frac{x_2^{i_2}}{{i_2}^{m_2}}
 \frac{x_3^{i_3}}{{i_3}^{m_3}}
\;\; =  \;\;
\begin{picture}(40,60)(-10,30)
\Vertex(10,50){2}
\Vertex(10,30){2}
\Vertex(10,10){2}
\Line(10,10)(10,50)
\Text(6,50)[r]{$x_1$}
\Text(6,30)[r]{$x_2$}
\Text(6,10)[r]{$x_3$}
\end{picture} 
\eq
\\
\\
The pictorial representation views a $Z$-sum as a rooted tree without any sidebranchings.
The outermost sum corresponds to the root. By convention, the root is always drawn
on the top.\\
\\
Trees with sidebranchings are given by nested sums with more than one subsum, for example:
\\
\vspace*{-15mm}
\bq
\sum\limits_{i=1}^n \frac{x_1^i}{i^{m_1}} Z(i-1;m_2,x_2) Z(i-1;m_3;x_3) 
& = &
\vspace*{-8mm}
\begin{picture}(60,60)(0,30)
\Vertex(30,50){2}
\Vertex(10,20){2}
\Vertex(50,20){2}
\Line(30,50)(10,20)
\Line(30,50)(50,20)
\Text(26,50)[r]{$x_1$}
\Text(6,20)[r]{$x_2$}
\Text(46,20)[r]{$x_3$}
\end{picture}
\eq
\\
Of course, due to the multiplication formula, trees with sidebranchings can always be
reduced to trees without any sidebranchings:
\bq
\lefteqn{
\sum\limits_{i=1}^n \frac{x_1^i}{i^{m_1}} Z(i-1;m_2,x_2) Z(i-1;m_3;x_3) 
 =  } & & \nonumber \\
 & &
Z(n;m_1,m_2,m_3;x_1,x_2,x_3) + Z(n;m_1,m_3,m_2;x_1,x_3,x_2) \nonumber \\
 & & 
 + Z(n;m_1,m_2+m_3;x_1,x_2x_3).
\eq
The coalgebra structure is formulated in terms of rooted trees (e.g. there is no need
to convert rooted trees to a basis of rooted trees without sidebranchings).
We first introduce some notation how to manipulate rooted trees.
We adopt the notation of Kreimer and Connes \cite{Kreimer:1998dp, Connes:1998qv}.
An elementary cut of a rooted tree is a cut at a single chosen edge. An admissible cut is any assignment of elementary
cuts to a rooted tree such that any path from any vertex of the tree to the root has at most one elementary cut.
An admissible cut maps a tree $t$ to a monomial in trees 
$t_1 \circ ... \circ t_{k+1}$. 
Note that precisely one of 
these subtrees $t_j$
will contain the root of $t$. We denote this distinguished tree by $R^C(t)$, and the monomial delivered by the $k$ other factors
by $P^C(t)$. The counit $\bar{e}$ is given by
\bq
\bar{e}(e) & = & 1, \nonumber \\
\bar{e}(t) & = & 0, \;\;\; t \neq e.
\eq
The coproduct $\Delta$ is defined by the equations
\bq
\Delta(e) & = & e \otimes e, \nonumber \\
\Delta(t) & = & e \otimes t + t \otimes e + \sum\limits_{\mbox{\tiny adm. cuts $C$ of $t$}} P^C(t) \otimes R^C(t), \nonumber \\
\Delta(t_1 \circ ... \circ t_k ) & = & \Delta(t_1) ( \circ \otimes \circ ) ... ( \circ \otimes \circ ) \Delta(t_k).
\eq
The antipode ${\cal S}$ is given by
\bq
{\cal S}(e) & = & e, \nonumber \\
{\cal S}(t) & = & -t - \sum\limits_{\mbox{\tiny adm. cuts $C$ of $t$}} {\cal S}\left( P^C(t) \right) \circ R^C(t), \nonumber \\
{\cal S}(t_1 \circ ... \circ t_k) & = & {\cal S}(t_1) \circ ... \circ {\cal S}(t_k).
\eq
The proof that these definitions yield a Hopf algebra can be found in \cite{Hoffman} and is not repeated here.
The Hopf algebra of 
Kreimer and Connes \cite{Kreimer:1998dp, Connes:1998qv},
which emerged in the context of renormalization of UV-divergences, has a slightly 
different algebra structure.
There the algebra is generated by rooted trees.
In this algebra a product of two rooted trees is not necessarily a rooted tree again.
However, the coalgebra structures are identical, which is a remarkable observation.\\
\\
Let us give some examples for the coproduct and the antipode for $Z$-sums:
\bq
\Delta Z(n;m_1;x_1) & = & 
 e \otimes Z(n;m_1;x_1) + Z(n;m_1;x_1) \otimes e, \nonumber \\
\Delta Z(n;m_1,m_2;x_1,x_2) & = &
 e \otimes Z(n;m_1,m_2;x_1,x_2) + Z(n;m_1,m_2;x_1,x_2) \otimes e \nonumber \\ 
 & & 
 + Z(n;m_2;x_2) \otimes Z(n;m_1;x_1),
\eq
\bq
{\cal S} Z(n;m_1;x_1) & = & 
 - Z(n;m_1;x_1), \nonumber \\
{\cal S} Z(n;m_1,m_2;x_1,x_2) & = &
 Z(n;m_2,m_1;x_2,x_1) + Z(n;m_1+m_2;x_1 x_2).
\eq

\section{Review of Goncharov's multiple polylogarithms}
\label{sec:polylog}

At the end of the day we express our results in terms of Goncharov's
multiple polylogarithms. They form therefore an important specialization
of nested sums and we review therefore some of their properties.
After the introduction by Gonacharov \cite{Goncharov}
they have been extensively studied by Borwein et al. \cite{Borwein}.
They use a different notation which is related
to the one of Goncharov by
\bq
\label{bnotation}
\mbox{Li}_{m_k,...,m_1}(x_k,...,x_1) = 
\lambda\left(
\begin{array}{c}
m_1,...,m_k \\
b_1,...,b_k \\
\end{array}
\right),
& & b_j = \frac{1}{x_1 x_2 ... x_j}.
\eq
Most of the material reviewed in this section is based on the work
of Borwein et al. \cite{Borwein}.

\subsection{Integral representations}

We first define the notation for iterated integrals
\bq
\int\limits_0^\Lambda \frac{dt}{a_n-t} \circ ... \circ \frac{dt}{a_1-t} & = & 
\int\limits_0^\Lambda \frac{dt_n}{a_n-t_n} \int\limits_0^{t_n} \frac{dt_{n-1}}{a_{n-1}-t_{n-1}} \times ... \times \int\limits_0^{t_2} \frac{dt_1}{a_1-t_1}.
\eq
We further use the following short hand notation:
\bq
\int\limits_0^\Lambda \left( \frac{dt}{t} \circ \right)^{m} \frac{dt}{a-t}
& = & 
\int\limits_0^\Lambda 
\underbrace{\frac{dt}{t} \circ ... \frac{dt}{t}}_{m \;\mbox{times}} \circ \frac{dt}{a-t}.
\eq
The integral representation for $\mbox{Li}_{m_k,...,m_1}(x_k,...,x_1)$ reads:
\bq
\label{intrepI}
\mbox{Li}_{m_k,...,m_1}(x_k,...,x_1) & = & \int\limits_0^{x_1 x_2 ... x_k} \left( \frac{dt}{t} \circ \right)^{m_1-1} \frac{dt}{x_2 x_3 ... x_k -t} \nonumber \\
& & \circ \left( \frac{dt}{t} \circ \right)^{m_2-1} \frac{dt}{x_3 ... x_k -t} \circ ... \circ
\left( \frac{dt}{t} \circ \right)^{m_k-1} \frac{dt}{1 -t}.
\eq
In the notation of Borwein et al. this representation reads
\bq
\label{intrepII}
\mbox{Li}_{m_k,...,m_1}(x_k,...,x_1) & = &
 (-1)^k \int\limits_0^1 \left( \frac{dt}{t} \circ \right)^{m_1-1} \frac{dt}{t-b_1} \nonumber \\
 & & 
 \circ \left( \frac{dt}{t} \circ \right)^{m_2-1} \frac{dt}{t-b_2}
 \circ ... \circ
 \left( \frac{dt}{t} \circ \right)^{m_k-1} \frac{dt}{t-b_k},
\eq
where the $b_j$'s are related to the $x_j$'s as in eq. (\ref{bnotation}).
Changing the integration variables according to $t \rightarrow 1-t$ yields
the dual integral representation:
\bq
\label{dualint}
\mbox{Li}_{m_k,...,m_1}(x_k,...,x_1) & = & (-1)^k
 \int\limits_0^1 \frac{dt}{1-b_k-t} 
   \circ \left( \frac{dt}{1-t} \circ \right)^{m_k-1} \nonumber \\
 & & 
 \frac{dt}{1-b_{k-1}-t} 
   \circ \left( \frac{dt}{1-t} \circ \right)^{m_{k-1}-1} 
 \circ ... \circ
 \frac{dt}{1-b_{1}-t} 
   \circ \left( \frac{dt}{1-t} \circ \right)^{m_{1}-1}. \nonumber \\
\eq
In addition to these weight-dimensional integral representations
there is also a
depth-dimensional integral representation:
\bq
\mbox{Li}_{m_k,...,m_1}(x_k,...,x_1) & = & \frac{1}{\Gamma(m_1) ... \Gamma(m_k)}
 \int\limits_1^\infty \frac{dt_1}{t_1} 
  \frac{ \left( \ln t_1 \right)^{m_1-1}}{ \frac{t_1}{x_1}-1} \nonumber \\
& & \times
 \int\limits_1^\infty \frac{dt_2}{t_2} 
  \frac{ \left( \ln t_2 \right)^{m_2-1}}{ \frac{t_1 t_2}{x_1 x_2}-1}
 \times ... \times
 \int\limits_1^\infty \frac{dt_k}{t_k} 
  \frac{ \left( \ln t_k \right)^{m_k-1}}{ \frac{t_1...t_k}{x_1...x_k}-1}.
\eq

\subsection{The shuffle algebra}

Instead of specifying a multiple polylogarithm by the $x_j$'s and $m_j$'s, we may denote
the function by a single string
\bq
\label{alphanotation}
\left( \alpha_1, \alpha_2, ..., \alpha_w \right)
 & = & 
\left( 0,..., 0, b_1, 0,..., 0, b_2, ..., 0, ..., 0, b_k \right),
\eq
where $(m_j-1)$ zeros preceed $b_j$. Defining further
\bq
\Omega(\alpha_i) & = & \frac{dt}{t-\alpha_i}
\eq
allows us to rewrite the integral representation eq. (\ref{intrepII}) as
\bq
\mbox{Li}_{m_k,...,m_1}(x_k,...,x_1) & = &
 (-1)^k \int\limits_0^1 
 \Omega(\alpha_1) \circ ... \circ \Omega(\alpha_w).
\eq
{}From the iterated integral representation one deduces a second algebra structure with
multiplication given by
\bq
\lefteqn{
\mbox{Li}_{m_k,...,m_1}(x_k,...,x_1) \times \mbox{Li}_{m_{k+l},...,m_{k+1}}(x_{k+l},...,x_{k+1}) } \nonumber \\
& = &
(-1)^{k+l} 
 \int\limits_0^1 \Omega(\alpha_1) \circ ... \circ \Omega(\alpha_{w_k})
 \int\limits_0^1 \Omega(\alpha_{w_k+1}) \circ ... \circ \Omega(\alpha_{w_k+w_l}) \nonumber \\
& = & (-1)^{k+l} \sum\limits_{shuffle} 
 \int\limits_0^1 \Omega(\alpha_{\sigma(1)}) \circ ... \circ \Omega(\alpha_{\sigma(w_k+w_l)}),
\eq
where $w_k=m_1+...+m_k$, $w_l=m_{k+1}+...+m_{k+l}$ and
the sum is over all permutations, which preserve the relative order of the strings
$\Omega(\alpha_1) ... \Omega(\alpha_{w_k})$ and $\Omega(\alpha_{w_{k}+1}) ... \Omega(\alpha_{w_k+w_l})$.

\subsection{Reduction to simpler functions}

The multiple polylogarithms contain a variety of other functions as a subset.
We start with depth one. As the notation already suggests, the multiple polylogarithms
are in this case identical to the classical polylogarithms, e.g.
\bq
\mbox{Li}_0(x) = \frac{x}{1-x},
& &
\mbox{Li}_1(x) = - \ln(1-x)
\eq
and
\bq
\mbox{Li}_n(x) = \int\limits_0^x dt \frac{\mbox{Li}_{n-1}(t)}{t}.
\eq
Nielsen's generalized polylogarithms \cite{Nielsen}, defined
through
\bq
S_{n,p}(x) & = & \frac{(-1)^{n-1+p}}{(n-1)! p!}
\int\limits_0^1 dt \frac{\ln^{n-1}(t) \ln^p(1-tx)}{t}
\eq
are related to the multiple polylogarithms by
\bq
S_{n,p}(x) & = & \mbox{Li}_{1,...,1,n+1}(\underbrace{1,...,1}_{p-1},x),
\eq
where $(p-1)$ one's occur before $n+1$ and $x$.
The harmonic polylogarithms of Remiddi and Vermaseren \cite{Remiddi:1999ew} are related
to the multiple polylogarithms for positive indices as
\bq
\label{HP}
H_{m_1,...,m_k}(x) & = & \mbox{Li}_{m_k,...,m_1}(\underbrace{1,...,1}_{k-1},x).
\eq
The harmonic polylogarithms are defined recursively through
\bq
H_0(x) = \ln(x), &\qquad  H_1(x) = - \ln(1-x), \qquad & H_{-1}(x) = \ln(1+x)\ ,
\eq
and
\bq
H_{m_1+1,m_2,...,m_k} & = & \int\limits_0^x dt f_0(t) H_{m_1,m_2,...,m_k}(t), \nonumber \\
H_{\pm 1,m_2,...,m_k} & = & \int\limits_0^x dt f_{\pm 1}(t) H_{m_2,...,m_k}(t),
\eq
where the fractions 
$f_0(x)$, $f_1(x)$ and $f_{-1}(x)$
are given by
\bq
f_0(x) = \frac{1}{x}, &\qquad  \displaystyle f_1(x) = \frac{1}{1-x}, \qquad & f_{-1}(x) = \frac{1}{1+x} .
\eq
Recently Gehrmann and Remiddi \cite{Gehrmann:2000zt}
extended the harmonic polylogarithms to two-dimensional
harmonic polylogarithms (2dHPL) by extending the fractions to
\bq
f(z,x) = \frac{1}{z+x}, & &
f(1-z,x) = \frac{1}{1-z-x}.
\eq
{}From the integral representation eq. (\ref{intrepI}) it is clear that the 2dHPL  
are a subset of Goncharov's multiple polylogarithms.
If we identitfy $x=x_1 x_2 \times ... \times x_k$ we have
\bq
\int\limits_0^x dt f(z,t) \mbox{Li}_{m_k,...,m_1}\left(x_k,...,x_2,\frac{t}{x_2...x_k}\right)
& = & 
- \mbox{Li}_{m_k,...,m_1,1}\left(x_k,...,x_2,-\frac{z}{x_2...x_k},-\frac{x}{z}\right),
\nonumber \\
\int\limits_0^x dt f(1-z,t) \mbox{Li}_{m_k,...,m_1}\left(x_k,...,x_2,\frac{t}{x_2...x_k}\right)
& = & 
\mbox{Li}_{m_k,...,m_1,1}\left(x_k,...,x_2,\frac{1-z}{x_2...x_k},\frac{x}{1-z}\right).
\nonumber \\
\eq

\end{appendix}

\end{document}